\documentclass[twocolumn,showpacs,
aps,superscriptaddress,
prd,notitlepage,showkeys,
nofootinbib, floatfix]{revtex4-1}

\usepackage[normalem]{ulem}
\usepackage{amssymb}
\usepackage{amsmath}
\usepackage{graphicx}
\usepackage{dcolumn}
\usepackage[colorlinks,urlcolor=blue,citecolor=blue,linkcolor=blue]{hyperref}
\usepackage{color,units}
\usepackage[dvipsnames]{xcolor} 
\usepackage{lineno}
\usepackage{xspace}
\usepackage{longtable} 
\usepackage{float} 
\usepackage{multirow}
\usepackage{amsfonts,wasysym,epsfig,verbatim,subfigure,bm,mathrsfs,lipsum}

\newcommand{\orcidicon}[1]{\href{https://orcid.org/#1}{\includegraphics[height=\fontcharht\font`\B]{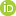}}}

\begin{document}

\newcommand{\IUCAA}{Inter-University Centre for Astronomy and
Astrophysics, Post Bag 4, Ganeshkhind, Pune 411 007, India}
\newcommand{\OKC}{The Oskar Klein Centre, Department of Astronomy, Stockholm University, AlbaNova, SE-10691 Stockholm,
Sweden}
\newcommand{\WSU}{Department of Physics \& Astronomy, Washington State University, 1245 Webster, Pullman, WA 99164-2814, USA}

\title{Simultaneous Inference of Neutron Star Equation of State and the Hubble Constant with a Population of Merging Neutron Stars}

\author{Tathagata Ghosh~\orcidicon{0000-0001-9848-9905}}
\affiliation{\IUCAA}

\author{Bhaskar Biswas~\orcidicon{0000-0003-2131-1476}}
\affiliation{\IUCAA}\affiliation{\OKC}

\author{Sukanta Bose~\orcidicon{0000-0002-4151-1347}}
\affiliation{\IUCAA}\affiliation{\WSU}

\date{\today}

\begin{abstract}

We develop a method for implementing a proposal on utilizing knowledge of neutron star (NS) equation of state (EoS) for inferring the Hubble constant from a population of binary neutron star (BNS) mergers.
This method is useful in exploiting BNSs as standard sirens when their redshifts are not available.
Gravitational wave (GW) signals from compact object binaries provide a direct measurement of their luminosity distances, but not their redshifts.
Unlike in the past, here we employ a realistic EoS parametrization in a Bayesian framework to simultaneously measure the Hubble constant and refine the constraints on the EoS parameters. 
The uncertainty in the redshift depends on the uncertainties in the EoS and the mass parameters estimated from GW data.
Combining the inferred BNS redshifts with the corresponding luminosity distances, one constructs a redshift-distance relation and deduces the Hubble constant from it.
Here, we show that in the Cosmic Explorer era, one can measure the Hubble constant to a precision of 
$\lesssim 5\%$ (with a $90\%$ credible interval)
with a realistic distribution of a thousand BNSs, while allowing for uncertainties in their EoS parameters.
Such a measurement can potentially resolve the current tension in the measurements of the Hubble constant from the early- and late-time universe.
The methodology implemented in this work demonstrates a comprehensive 
prescription for inferring the NS EoS and the Hubble constant by simultaneously combining GW observations from merging NSs, while employing a simple population model for NS masses and keeping the merger rate of NSs constant in redshift.
This method can be immediately extended to incorporate merger rate, population properties, and additional cosmological parameters.

\end{abstract}

\maketitle

\section{Introduction}

Gravitational wave (GW) observations of compact binary coalescences (CBC) have opened a new era of precision cosmology. One of the key goals is to measure the Hubble constant (denoted by $H_{0}$), the current expansion rate of the Universe.
GW allows the direct measurement of luminosity distance. Additionally, if the redshifts of GW sources are available from any other observations, the Hubble constant can be estimated from the distance-redshift relation. 
In the absence of an electromagnetic (EM) counterpart, the Hubble constant can be measured 
using statistical correlation~\cite{Schutz:1986gp, Chen:2017rfc, LIGOScientific:2018gmd, Nair:2018ign, Gray:2019ksv, Borhanian:2020vyr, Bera:2020jhx, Mukherjee:2020hyn} of distance with an ensemble of potential host galaxies within the GW event localization uncertainty region. 
The first ever multimessenger detection of binary neutron star (BNS) merger  GW170817~\cite{LIGOScientific:2017vwq} by the Advanced LIGO \cite{LIGOScientific:2014pky} and Advanced Virgo detectors \cite{VIRGO:2014yos}, with unique identification of host galaxy NGC 4993, is the first standard siren used for the measurement of the Hubble constant $H_0=70^{+12}_{-8}\ \rm{km\ s}^{-1}\ \rm{Mpc}^{-1}$~\cite{LIGOScientific:2017adf}.
However, it is not clear what fraction of BNSs observed in GWs may have EM counterparts that yield a precise redshift measurement, especially at large distances. In this work, we make the conservative assumption that this fraction is small~\cite{DES:2017kbs, LIGOScientific:2017adf}. 

Apart from bright sirens, like GW170817, $H_{0}$ can also be estimated from dark sirens, which lack confirmed EM counterparts.
The LIGO-Virgo-KAGRA collaboration recently published the updated measurements of the Hubble constant~\cite{LIGOScientific:2021aug} using $47$ GW events from GWTC-3~\cite{LIGOScientific:2021djp} to be $68^{+12}_{-8}\ \rm{km\ s}^{-1}\ \rm{Mpc}^{-1}$ and $68^{+8}_{-6}\ \rm{km\ s}^{-1}\ \rm{Mpc}^{-1}$ following two different methods, as described in Ref.~\cite{Mastrogiovanni:2021wsd} and Ref.~\cite{Gray:2019ksv}, respectively.
These recent measurements are clearly an improvement compared to the previous estimate of $H_{0}=69^{+16}_{-8}\ \rm{km\ s}^{-1}\ \rm{Mpc}^{-1}$ -- from observations during the O1-O2 runs~\cite{LIGOScientific:2019zcs}.
Unsurprisingly, all of these GW constraints on $H_{0}$ have significant contribution from the observations of GW170817 and its EM counterpart.
More instances of independent measurements of luminosity distance and redshift from a wide distribution of sources would be useful toward a resolution of the current tension in $H_{0}$ measurements from the early and late-time universe \cite{Verde:2019ivm}.
To wit, there is a $4.4\sigma$ discrepancy between the values 
of $H_{0}$ as measured from the cosmic microwave background ($67.36 \pm 0.54\ \rm{km\ s}^{-1}\ \rm{Mpc}^{-1}$~\cite{Planck:2018vyg}) and observations based on the cosmic distance ladder comprising Cepheid variable stars and supernovae type Ia ($73.04 \pm 1.04\ \rm{km\ s}^{-1}\ \rm{Mpc}^{-1}$~\cite{Riess:2021jrx}).

Messenger and Read \cite{Messenger:2011gi} proposed an alternative idea of determining the redshift of a binary neutron star (BNS) merger in a GW observation by using prior knowledge of the neutron star (NS) equation of state (EoS). 
The phase $\Psi(f)$ of the frequency domain GW waveform, $\tilde{h}(f)=A(f)e^{-i\Psi(f)}$ (where $A(f)$ is the amplitude of GW waveform at frequency $f$), has two components: one of them is the standard post-Newtonian point-particle frequency domain phase
$\Psi_{\text{PP}}(f)$ and the other is the  phase component $\Psi_{\text{tidal}}(f)$ due to the tidal deformability of the neutron star, i.e., $\Psi(f)=\Psi_{\text{PP}}(f)+\Psi_{\text{tidal}}(f)$. 
$\Psi_{\text{PP}}(f)$ depends on the \textit{redshifted} chirp mass $\mathcal{M}_{c}^{z}=(1+z)\mathcal{M}_{c}$ and the redshifted frequency $f = f_{\text{source}}/(1+z)$ in such a way that 
the redshift factor $(1+z)$ cancels out of it, leaving it  redshift invariant~\cite{Messenger:2011gi}.
Here,  $\mathcal{M}_{c}$ and $f_{\text{source}}$ are the source-frame chirp mass and source-frame GW frequency, respectively.
However, $\Psi_{\text{tidal}}$ contains NS tidal deformability terms~\cite{Flanagan:2007ix}, which depend on the source-frame (i.e., unredshifted) masses.
The degeneracy between the mass parameters and the redshift can be broken if one knows the EoS of NSs~\cite{Messenger:2011gi} 
since its precise knowledge can be used to determine the tidal deformability uniquely for a given source-frame mass of NS.
This technique is especially useful when the BNS is not accompanied by an electromagnetic counterpart to infer redshift directly. 
This method neither depends on the observation of EM counterpart nor the identification of potential host galaxies for statistical cross-correlation methods.

Recent progress was reported by Chatterjee \textit{et al.}~\cite{Chatterjee:2021xrm}
by building on the basic concept proposed in Ref.~\cite{Messenger:2011gi}.
Their key idea was to use binary Love relations~\cite{Yagi:2015pkc} to capture information about NS EoS in a single tidal parameter and use it to constrain the Hubble constant. 
In this work, we too follow Ref.~\cite{Messenger:2011gi} but use a hybrid nuclear plus piecewise-polytrope (PP) parametrization to model the NS EoS~\cite{Biswas:2020puz, Biswas:2021yge}.
Specifically, at low densities in the star the nuclear empirical parameters are used to construct the EoS model. These nuclear empirical parameters (e.g., nuclear symmetry energy) can be constrained from laboratory experiments and astrophysical observation of neutron stars.
The NS EoS, however, is less well understood at densities that are about a couple of times the nuclear saturation density or higher.
In those regions, we use the three-piece PP model. 
We will briefly discuss the EoS model in the next section.

The structure of the paper is as follows. Section~\ref{EoS_model} introduces the construction of the EoS model. In Sec.~\ref{method}, we describe the Bayesian framework to infer the Hubble constant.
In Secs.~\ref{simulation} and ~\ref{results}, we discuss the required simulations and the results, respectively.
For our simulations, we take the $\Lambda$CDM cosmology with $\Omega_{m}=0.3$ and $H_{0}= 70\ \rm{km\ s}^{-1}\ \rm{Mpc}^{-1}$ as the \textit{true} cosmology throughout.

\section{EoS model: Hybrid Nuclear+PP } \label{EoS_model}

In this section, we briefly introduce hybrid nuclear+PP EoS parametrization, which has been used in the past~\cite{Biswas:2020puz,Biswas:2021yge} to put joint GW-electromagnetic constraints on the properties of neutron stars. This hybrid EoS parametrization is also used to investigate the nature of the ``mass-gap'' object in GW190814~\cite{Biswas:2020xna}. Since the crust has 
minimal impact~\cite{Gamba:2019kwu,Biswas:2019ifs} on the macroscopic properties of NSs such mass, radius, tidal deformability, etc., standard BPS EoS~\cite{1971ApJ...170..299B} is used to model the crust under this parametrization. Then this fixed crust is joined with the core EoS in a thermodynamically consistent fashion described in Ref.~\cite{Xie:2019sqb}. The core EoS is divided into two parts: 

1. EoS around the nuclear saturation density ($\rho_0$) is well described via the parabolic expansion of energy per nucleon
\begin{equation}
    e(\rho,\delta) \approx  e_0(\rho) +  e_{\rm sym}(\rho)\delta^2,
\end{equation}
where $e_0(\rho)$ is the energy of symmetric nuclear matter for which the number of protons is equal to the number of neutrons, $e_{\rm sym}$ is the energy of the asymmetric nuclear matter (commonly referred as ``symmetry energy'' in literature), and $\delta \equiv \frac{\rho_p-\rho_n}{\rho_p+\rho_n}$ (where $\rho_n,\rho_p$ being the number density of neutron and proton respectively) is the measure of asymmetry in the number density of neutrons and protons.  Around $\rho_0$, these two energies can be further expanded in a Taylor series: 
\begin{eqnarray}
 e_0(\rho) &=&  e_0(\rho_0) + \frac{ K_0}{2}\chi^2 \label{eq:e0} +\,...,\\
e_{\rm sym}(\rho) &=&  e_{\rm sym}(\rho_0) + L\chi + \frac{ K_{\rm sym}}{2}\chi^2 
 ..., \label{eq:esym}
\end{eqnarray}
where $\chi \equiv (\rho-\rho_0)/3\rho_0 \ll 1$. We truncate the Taylor expansion up to the second order in $\chi$ as we use this expansion only up to $1.25 \rho_0$. The lowest order parameters are experimentally well constrained and therefore, we fix them at their median values such as $e_0(\rho_0) = -15.9$ MeV, and $\rho_0 =0.16 \rm{fm^{-3}}$. Therefore, the free parameters of this nuclear-physics informed  model are the curvature of symmetric matter $K_{0}$, nuclear symmetry energy $e_{sym}$, slope $L$, and curvature of symmetric energy $K_{sym}$. A survey based on 53 experimental results performed in 2016~\cite{Oertel:2016bki} found of values of $e_{\rm sym} (\rho_0) = 31.7 \pm 3.2 $ MeV and $L = 58.7 \pm 28.1$ MeV. 
Using these values as a prior, a Bayesian analysis performed in Ref.~\cite{Biswas:2020puz} combining multiple astrophysical observations (GWs and x-rays) has already provided a better constraint on these quantities: $e_{\rm sym} (\rho_0) = 32.0^{+3.05}_{-3.01} $ MeV and $L = 61.0^{+17.7}_{-16.0}$ MeV.

2. At higher densities, i.e., above $1.25 \rho_0$, the empirical parametrization starts to break down. Therefore, following Ref.~\cite{Read:2008iy}, at such densities we use a three-piece piecewise-polytrope  parametrization ($\Gamma_1,\Gamma_2, \rm{and}\ \Gamma_3$ being the polytropic indices),
divided by the fixed transition densities at $10^{14.7} \mathrm{gm/cm^3}$ and $10^{15} \mathrm{gm/cm^3}$ respectively.    

Instead of using all the EoS parameters explicitly, we will generally use $\bm{\mathcal{E}}=\{ K_{0}, e_{sym}, L, K_{sym} , \Gamma_{1},\Gamma_{2},\Gamma_{3} \}$. 
In our work, we consider this particular EoS model to constrain the cosmological parameters. 
In future, different choices of EoS parametrization can be explored.
One such example is the newly developed EoS-insensitive approach~\cite{Biswas:2021paf}  that has considerably less systematic error compared to some other EOS-insensitive approaches~\cite{Yagi:2015pkc, Yagi:2016qmr} available in the literature.

\section{Methodology} \label{method}

\begin{figure*}
    \centering
    \includegraphics [width = 17cm,height=6cm] {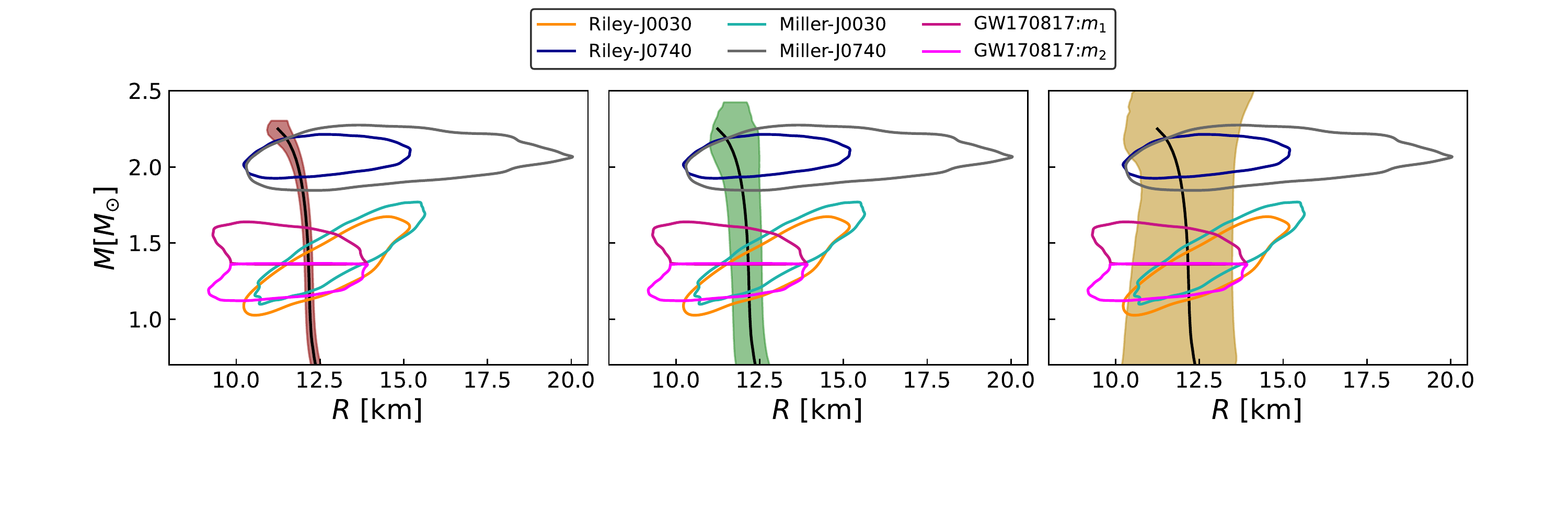}
    \caption{$90\%$ credible interval of mass-radius priors are shown  corresponding to three different choices of EoS prior distributions, as described in Table \ref{tab:EoS_prior}. The solid black line corresponds to the injected EoS.}
    \label{fig:eos_prior}
\end{figure*}

\begingroup
\renewcommand{\arraystretch}{1.2}
\begin{table*}
\caption{\label{tab:EoS_prior}
Different choices of uncertainties in EoS parameters that we use to establish the effectiveness of our method, as described in Sec.~\ref{EoS_model}. Here $\mathcal{N}(\mu, \sigma)$ refers to the Gaussian distribution with mean $\mu$ and standard deviation $\sigma$.}
\begin{ruledtabular}
\begin{tabular} {ccccc}
EoS Parameters & Injected & Prior 1 & Prior 2 & Prior 3 \\
\colrule
$K_{0}$ [MeV] & $240$ & $\mathcal{N}(240,2.5)$ & $\mathcal{N}(240,20)$ & $\mathcal{N}(240,30)$\\
$e_{\rm{sym}}$ [MeV] & $31.7$ & $\mathcal{N}(31.7,0.5)$ & $\mathcal{N}(31.7,2.5)$ & $\mathcal{N}(31.7,3.2)$\\
$L$ [MeV] & $58.7$ & $\mathcal{N}(58.7,1.5)$ & $\mathcal{N}(58.7,5)$ & $\mathcal{N}(58.7,28.1)$\\
$K_{\rm{sym}}$ [MeV] & $-100$ & $\mathcal{N}(-100,2.5)$ & $\mathcal{N}(-100,50)$ & $\mathcal{N}(-100, 100)$\\
$\Gamma_{1}$ & $2.5$ & $\mathcal{N}(2.5,0.05)$ & $\mathcal{N}(2.5,0.2)$  & $\mathcal{N}(2.5,0.5)$\\
$\Gamma_{2}$ & $3.5$ & $\mathcal{N}(3.5,0.05)$ & $\mathcal{N}(3.5,0.2)$ & $\mathcal{N}(3.5,0.5)$\\
$\Gamma_{3}$ & $3.0$ & $\mathcal{N}(3.0,0.5)$ & $\mathcal{N}(3.0,2.0)$ & $\mathcal{N}(3.0,2.5)$\\
\end{tabular}
\end{ruledtabular}
\end{table*}
\endgroup

Gravitational-wave observations of the leading binary phase terms, in $\Psi_{\text{PP}}$,
provide mass estimation in the detector-frame; i.e., the source-frame mass $m$ is observed as redshifted mass, defined as $m^{z}=m(1+z)$, where $z$ is the redshift of the source.
Contrastingly, tidal deformability measurements of a BNS merger in GW observations depend on the source-frame mass pair.
Since the detector-frame masses are well estimated from post-Newtonian (PN) terms at lower orders, tidal deformability can provide the redshift information if we know the NS EoS~\cite{Messenger:2011gi}.
The redshift of the BNS, so obtained, can be combined with its luminosity distance to infer the Hubble constant.
However, it has proved quite challenging to measure the NS EoS very precisely. 
While it is fair to expect that it will be better constrained by the time of the third generation GW detectors, e.g., Cosmic Explorer (CE) \cite{Evans:2021gyd} and Einstein Telescope (ET) \cite{Maggiore:2019uih},
nevertheless, there will always be some uncertainty present in the values of the EoS parameters even in that era~\cite{Evans:2021gyd, Maggiore:2019uih}.
We outline a Bayesian framework for inferring the cosmological parameters from GW observations of BNSs with (imprecise) knowledge of NS EoS.
We allow for realistic uncertainties in the values of the EoS parameters and assign corresponding priors to them.
These priors are then utilized to infer $H_{0}$ and simultaneously refine the constraints on the NS EoS parameters.
Indeed, the joint posterior of the Hubble constant and the NS EoS parameters is given by
\begin{widetext}
\begin{eqnarray} \label{bayesH0}
    p (H_{0}, \bm{\mathcal{E}} \mid \{x_{\rm{GW_{i}}}\} ) \propto \prod_{i=1}^{N_{\rm{det}}} \frac{\pi(H_{0})}{\beta (H_{0})} \pi ( \bm{\mathcal{E}})\ && \iiint p \big(x_{\rm{GW}_{i}} \mid m_{1,i}^{z}(m_{1,i},z_{i}), m_{2,i}^{z}(m_{2,i},z_{i}), \Lambda_{1,i}(m_{1,i}, \bm{\mathcal{E}}), \Lambda_{2,i}(m_{2,i}, \bm{\mathcal{E}}), d_{L,i} (z_{i}, H_{0}) \big)\nonumber \\
    && \times \pi(m_{1,i}, m_{2,i} \mid \bm{\mathcal{E}} ) \pi(z_{i})  dz_{i} dm_{1,i} dm_{2,i}\,,  
\end{eqnarray}
\end{widetext}
where,
the subscript $i$ denotes the $i$th GW event.
In Eq.~\eqref{bayesH0}, $x_{\rm{GW}}$ is the GW data,
$m_{1,2}^{z}$ are the detector-frame (redshifted) masses of a BNS corresponding to the source-frame masses $m_{1,2}$ at redshift $z$, $\Lambda_{1,2}$ are the tidal deformabilities, and $d_{L}$ is the luminosity of the source (which is related to its $z$, via the Hubble constant $H_{0}$).
Here, $\pi(H_{0})$, $\pi(\bm{\mathcal{E}})$ and $\pi(z)$ are the prior distributions over the Hubble constant, EoS parameters and redshift, respectively.
$\pi(m_{1}, m_{2} \mid \bm{\mathcal{E}})$ is the prior distribution over the source-frame component masses determined from the NS population model and the choice of EoS parameters $\bm{\mathcal{E}}$.
$\beta({H_{0}})$ is a normalization factor, which ensures that the GW likelihood factor $p(x_{\rm{GW}}\mid m_{1}^{z}, m_{2}^{z}, \Lambda_{1}, \Lambda_{2}, d_{L})$ integrated 
over all observable sources is unity.

In our work, the mass distribution of NS is chosen to be uniform as follows:

\begin{equation} \label{mass_dist}
    p_{\text{pop}} (m\mid \bm{\mathcal{E}})= 
\begin{cases}
    \frac{1}{ m_{\text{max}}(\bm{\mathcal{E}}) - m_{\text{min}} },& \text{iff}\ \  m_{\text{min}}\leq m \leq m_{\text{max}} \\
    0,              & \text{otherwise\,,}
\end{cases}
\end{equation}
where $m_{\text{min}}$ is the minimum mass and $m_{\text{max}}$ is the maximum mass for the particular EoS parameters $\bm{\mathcal{E}}$.
In this work, we fix $m_{\text{min}}= 1M_{\odot}$, which is consistent with the predicted lower bound of NS mass from plausible supernova formation channels~\cite{2012ApJ...749...91F, Woosley:2020mze} and standard search of BNS merger greater than $1M_{\odot}$ by LIGO-Virgo~\cite{LIGOScientific:2016hpm}.
Specifically, in our population model the BNS mass distribution is taken as:

\begin{equation} \label{mass_bns}
\pi(m_{1}, m_{2} \mid \bm{\mathcal{E}}) \propto p_{\text{pop}} (m_{1} \mid \bm{\mathcal{E}}) p_{\text{pop}} (m_{2} \mid \bm{\mathcal{E}})\,,
\end{equation}
subject to the constraint $m_{1} \geq m_{2}$.

In Eq.~\eqref{bayesH0}, the prior over the redshift of the source is represented by $\pi(z)$, taken to be of the form 
\begin{eqnarray} \label{prior_redshift}
    \pi (z) \propto \frac{dV_{c}}{dz} \frac{R(z)}{1+z}\,.
\end{eqnarray}
Here, $V_{c}(z)$ is the comoving volume as a function of redshift, and $R(z)$ is the merger-rate density
in the detector-frame. 
The factor $(1+z)^{-1}$ above converts 
the merger rate from the source-frame time to the  detector-frame time.
In general, the merger rate density may be a function of redshift. 
In our work, we take $R(z)$=constant, i.e., the BNS distribution is considered to be uniform in comoving volume and source-frame time throughout this work.
The prior over the Hubble constant $\pi(H_{0})$ is chosen to be uniform. The prior ranges of all the relevant parameters are mentioned in Sec.~\ref{simulation}.

In our approach, we simultaneously estimate the EoS parameters and $H_{0}$ using a two-step Bayesian analysis.
In the first step, 
we directly measure, in the detector frame,
source parameters
like $m_{1,2}^{z}$, $d_{L}$, $\Lambda_{1,2}$  and inclination angle $\iota$ from the GW signals of individual BNSs using some prior $\pi (m_{1}^{z}, m_{2}^{z},\Lambda_{1}, \Lambda_{2}, d_{L}, \iota)$
over them. 
This step results in the construction of posteriors on those
parameters.
A detailed discussion on this aspect is presented in Sec.~\ref{simulation}. 
In the second step, these posterior distributions $p(m_{1}^{z}, m_{2}^{z}, \Lambda_{1}, \Lambda_{2}, d_{L}\mid  x_{\rm{GW}})$ of a single GW event, marginalized over the complementary set of signal parameters ($\iota$ in our work), are utilized to construct the GW likelihood factor $p(x_{\rm{GW}} \mid m_{1}^{z}, m_{2}^{z}, \Lambda_{1}, \Lambda_{2}, d_{L})$ by dividing out the \textit{prior} $ \pi (m_{1}^{z}, m_{2}^{z},\Lambda_{1}, \Lambda_{2}, d_{L})$ using the Bayes theorem:
\begin{eqnarray} \label{gwpost}
    p(m_{1}^{z}, m_{2}^{z}, \Lambda_{1}, \Lambda_{2}, d_{L}\mid  x_{\rm{GW}}) && \propto p(x_{\rm{GW}}\mid  m_{1}^{z}, m_{2}^{z}, \Lambda_{1}, \Lambda_{2}, d_{L}) \nonumber \\
    && \times \pi (m_{1}^{z}, m_{2}^{z},\Lambda_{1}, \Lambda_{2}, d_{L})\,,
\end{eqnarray}
The semimarginalized GW likelihood term is now used to produce the joint posterior of the EoS parameters and $H_{0}$ by employing Eq.~\eqref{bayesH0}.
Finally, we combine the posteriors of EoS parameters and $H_{0}$ of individual BNS events hierarchically, as elaborated in Appendix~\ref{bayesian_details}.

In Eq.~\eqref{bayesH0}, the normalization term $\beta(H_{0})$
encodes the selection effect~\cite{Mandel:2018mve, Chen:2017rfc}; 
it is defined as:
\begin{equation} \label{beta}
    \beta(H_{0}) = \int p_{\rm{det}} (z, H_{0}) \pi (z) dz\,,
\end{equation}
where $p_{\rm{det}}$ denotes the detection probability of any GW event. 
In Appendix~\ref{selection_effect}, we describe how $p_{\rm{det}}$ is computed in our simulations.

When quantifying the selection effect, we should ideally consider the effect of 
the uncertainty in the NS EoS parameters and the population of BNSs since
$p_{\rm{det}}$ depends on the properties of the GW source population.
The uncertainty in EoS parameters also affects the mass distribution, as maximum mass depends on the values of EoS parameters. 
The effect of tidal deformability on the amplitude of the gravitational wave signal is small~\cite{LIGOScientific:2019eut, Chatziioannou:2020pqz}. So, we have fixed the EoS parameters at the injection values and, hence, 
have also fixed the
maximum mass ($\approxeq 2.27 M_{\odot}$) to correspond 
to the injected EoS parameters. 
For simplicity, we assume that the population properties are known exactly.
However, it will be a good exercise to include the uncertainties in the properties of the GW source population in future work~\cite{dark_bns}.
We prepare the mock GW catalog assuming that the masses of BNSs follow a
uniform distribution in the source frame such that $m_{1} \geq m_{2}$
[see, Eq.~\eqref{mass_bns}].
It is clear from Eq.~\eqref{mass_dist} that the population of BNS depends on the minimum and maximum mass of BNS.
The minimum and maximum mass are fixed in the simulation.
However, the maximum mass is determined by the EoS of NS.
Since we do not account for the uncertainty of EoS parameters for estimating the selection effect, the maximum mass is also fixed (to the value mentioned above). 
So, the population of BNS does not affect the selection effect. 

We assumed flat $\Lambda$CDM cosmology with $H_{0}$ as a free parameter keeping $\Omega_{m}$ fixed at $0.3$,
while performing our Bayesian formulation, Eq.~\eqref{bayesH0}. 
In flat $\Lambda$CDM universe, the luminosity distance $d_{L}$ is related to redshift $z$ as 
\begin{equation} \label{redshift-dist}
    d_{L} (z) = \frac{c(1+z)}{H_{0}} \int_{0}^{z} \frac{dz'}{\sqrt{\Omega_{m}(1+z')^{3}+(1-\Omega_{m})}}\,.
\end{equation}
In the $3$rd generation detector era, it is expected to estimate $\Omega_{m}$ along with $H_{0}$, as the $3$rd generation detectors will probe the high redshift universe, resulting in the detection of a much large number of GW sources~\cite{Evans:2021gyd, Maggiore:2019uih}. 
It is also straightforward to incorporate  $\Omega_{m}$ as an unknown parameter in the Bayesian framework (Eq.~\eqref{bayesH0}). 
Since here we consider $50$ GW sources, between $100$ Mpc and $8$ Gpc, allowing $\Omega_{m}$ to vary would have a negligible impact on $H_{0}$ estimation with these few events~\cite{Mastrogiovanni:2021wsd}.

We have used \verb+PyMultinest+~\footnote{ \url{https://johannesbuchner.github.io/PyMultiNest/}}~\cite{Buchner:2014nha}, python based package based on the nested sampling algorithm Multinest to perform the parameter estimation within the Bayesian framework as mentioned in Eq.~\eqref{bayesH0}. During this Bayesian analysis, we use the multivariate Gaussian kernel density estimator from \verb+statsmodels+~\footnote{\url{https://www.statsmodels.org/}}~\cite{seabold2010statsmodels} to calculate the GW likelihood term.

\section{Simulation} \label{simulation}
\begin{figure*} 
    \centering
    \includegraphics[width = 16cm,height=16cm]{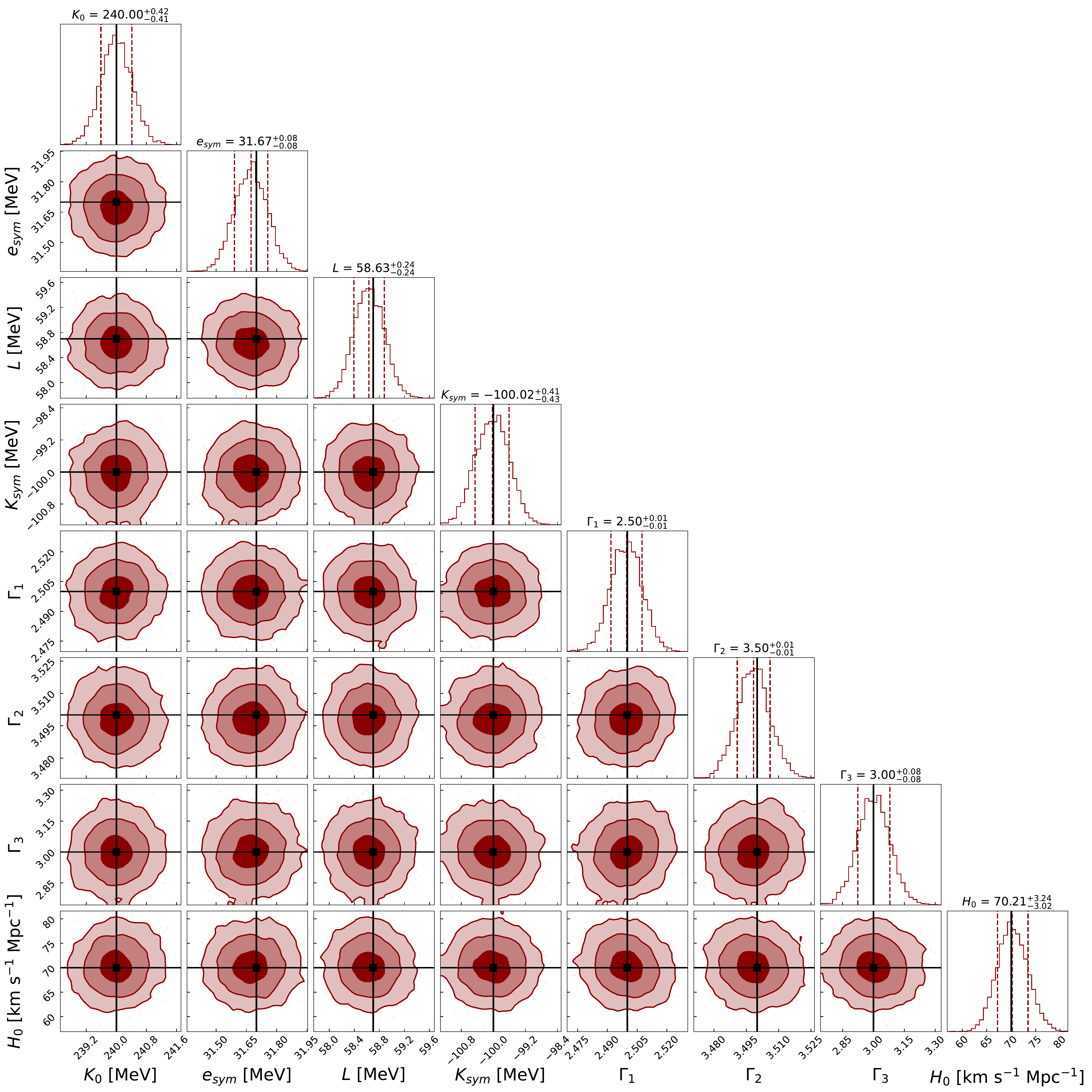}
    \caption{Constraint on the EoS parameters and $H_{0}$ by combining 50 GW events as detected by CE using EoS Prior 1. The black lines show the injected values. The median and $1 \sigma$ credible intervals are also shown in the respective marginalized one-dimensional posterior.}
    \label{fig:CE_EOS1}
\end{figure*}
\begin{figure*} 
    \centering
    \includegraphics[width = 16cm,height=16cm]{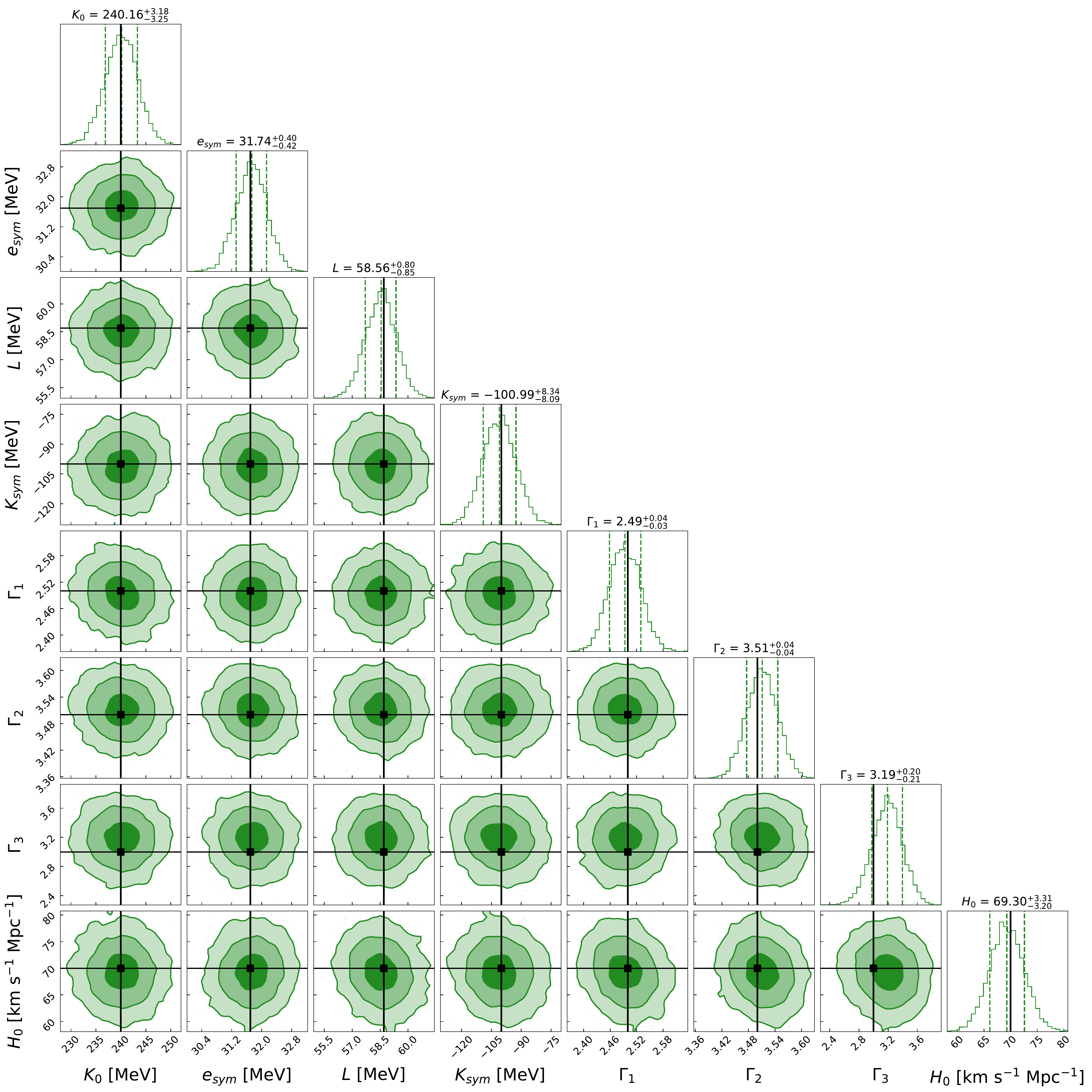} 
    \caption{Same as in Fig.~\ref{fig:CE_EOS1} but now using EoS Prior 2.}
    \label{fig:CE_EOS2}
\end{figure*}
\begin{figure*}
    \centering
    \includegraphics[width = 16cm,height=16cm]{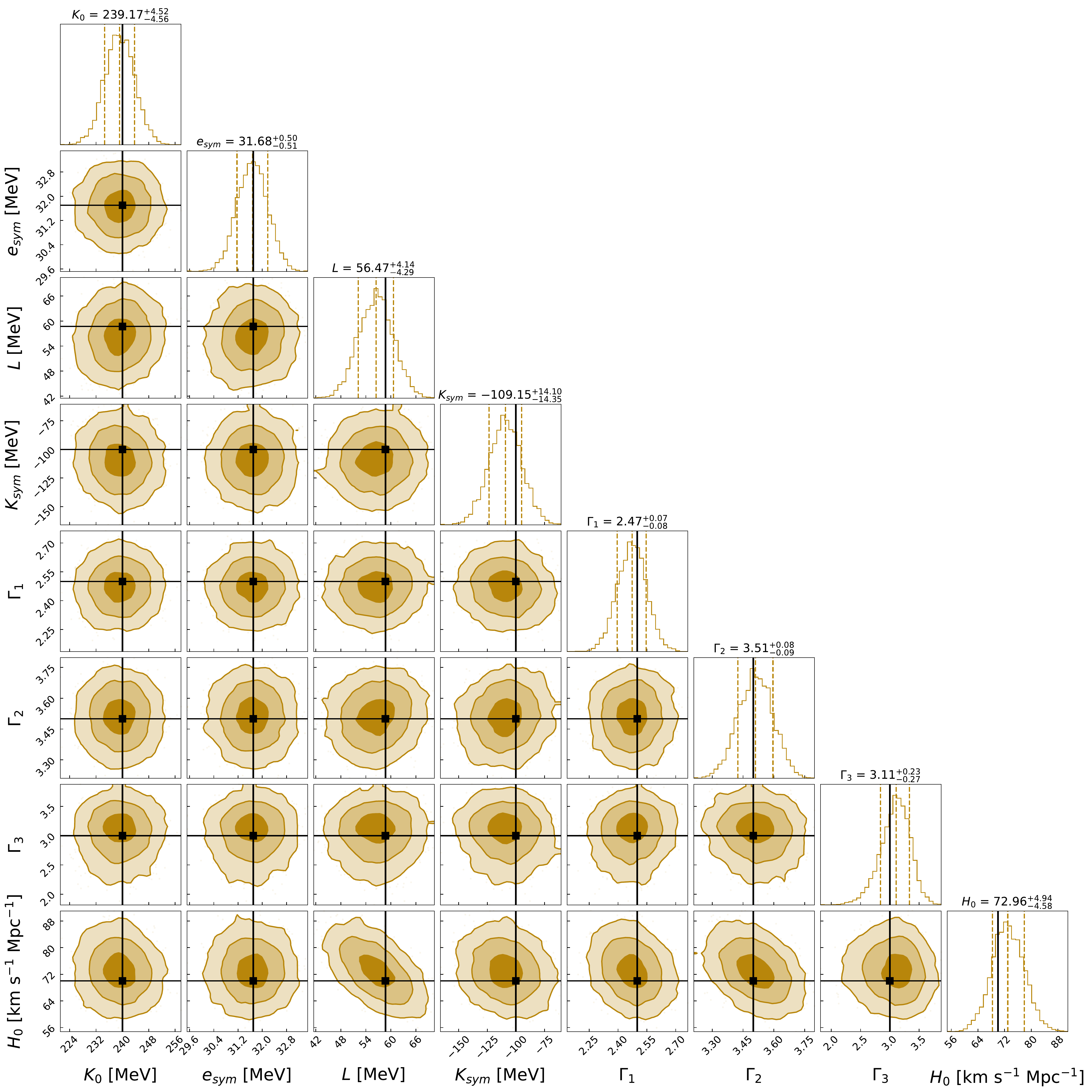} 
    \caption{Same as in Fig.~\ref{fig:CE_EOS1} but now using EoS Prior 3.}
    \label{fig:CE_EOS3}
\end{figure*}

We test the method described in Sec.~\ref{method} on several simulated GW events.
We first created a mock GW catalog by injecting the sources uniformly over the sky.
The redshift distribution of the sources follows Eq.~\eqref{prior_redshift}, with $R(z)$=constant and $d_{L}$ between $100$Mpc and $8$Gpc.
The masses of the BNS mergers follow uniform distribution in the source frame, as mentioned in Eq.~\eqref{mass_bns}, where $m_{\rm min}=1 M_{\odot}$ and $m_{\rm max}=2.27 M_{\odot}$. However, the uniform prior in source-frame mass is not a realistic mass distribution~\cite{LIGOScientific:2021psn}. We consider a uniform distribution over the source-frame mass for simplicity. 
However, it will be worth applying this methodology 
in the future
to different population models of BNS to understand the effect of the population on the simultaneous measurement of the Hubble constant and the EoS parameters.
(We briefly discuss the importance of 
employing realistic population models and merger rates in Sec.~\ref{discussion}.)
We consider flat $\Lambda$CDM cosmology with $\Omega_{m}=0.3$ and $H_{0}=70\ \rm{km\ s}^{-1}\ \rm{Mpc}^{-1}$ to determine the redshift of the GW sources from their luminosity distances. 
We ignore the spins of neutron stars (set to $0$) in the present work.

Tidal deformability of a neutron star depends on its mass, given a particular EoS. 
The injected values of the EoS parameters are $K_{0}=240$ MeV, $e_{\rm{sym}}=31.7$ MeV, $L=58.7$ MeV and $K_{\rm{sym}}=-100$ MeV, $\Gamma_{1} = 2.5$, $\Gamma_{2} = 3.5$ and $\Gamma_{3}=3.0$.
In Fig.~\ref{fig:eos_prior}, we show the injected EoS, which is consistent with the constraints on the mass-radius measurement from PSR J0030+0451~\cite{Riley:2019yda, Miller:2019cac} and PSR J0740+6620~\cite{Riley:2021pdl, Miller:2021qha} by the NICER collaboration; and GW170817 by LIGO-Virgo observations~\cite{LIGOScientific:2018cki}.
We consider CE as the detector in our simulation.
We inject 
a synthetic BNS signal of $128$ sec duration in stationary Gaussian noise with 
CE sensitivity~\footnote{\url{https://dcc.ligo.org/LIGO-T1500293/public}}~\cite{Abbott_2017}.
Though CE will observe a much longer signal, the tidal effect, which 
begins at $5$PN, is more pronounced closer to the merger \cite{Messenger:2011gi, Chatterjee:2021xrm}.
We consider the same waveform model \verb+IMRPhenomPv2_NRTidal+~\cite{Dietrich:2019kaq} for both 
signal injection and recovery.
We used \verb+bilby_pipe+~\footnote{\url{https://pypi.org/project/bilby-pipe/}}, a package for automating the process of running \verb+bilby+~\footnote{\url{https://lscsoft.docs.ligo.org/bilby/}}~\cite{Ashton_2019} for gravitational wave parameter estimation 
in computing clusters, to estimate the posterior of the parameters of BNS signals. We performed parameter estimation over the mass pair, tidal deformabilities, luminosity distance and inclination angle. For mass estimation, we consider uniform prior over observed chirp-mass  (in the range $\mathcal{M}_{c, \text{inj}}^{z} \pm 0.1 \ M_{\odot}$) and mass-ratio $q=m_{2}/m_{1} \in [0.2, 1]$. We use a uniform prior in tidal deformabilities $\Lambda_{1,2} \in [0,5000]$. 
For estimating the luminosity distance, we choose the prior 
in the comoving distance to be uniform and between $10$ Mpc and $2d_{L, \text{inj}}$.
We consider isotropic prior over to infer the inclination angle.
We fixed the remaining parameters (namely, sky-position, dimensionless spin magnitudes, tilt angle between their spins, orbital angular momentum, spin vectors describing the azimuthal angle separating the spin vectors, precession angle, polarization angle, coalescence time and its orbital phase at coalescence) 
at the injected values, although one should ideally use broad enough priors for all the source parameters.

In particular, fixing GW amplitude parameters affects the distance estimation and, hence, $H_{0}$.
Computational constraints restricted us in carrying out our analysis using only one detector (CE). 
Since it is not possible to measure the GW source's sky-position in such a scenario, 
we took it as known when estimating its other parameters. 
This is a limitation of this work that should be improved in the future.
In a realistic scenario, employing multiple detectors can help in
mitigating the luminosity distance-inclination angle degeneracy significantly.
So, it would be an interesting follow-up to this work to consider a more realistic situation with multiple detectors (for example, CE and ET, in the 3G era), allowing GW source parameters (especially GW amplitude parameters) for future investigation of the effects on the measurability of the EoS parameters and $H_{0}$.

Now, the parameters of the synthetic GW catalog have been used to determine $H_{0}$ and the EoS parameters, as discussed in Sec.~\ref{method}. 
We consider sufficiently large uniform prior over redshift to cover the entire luminosity distance posterior obtained from GW observation for any value of $H_{0}$ within our prior range between $10\ \rm{km\ s}^{-1}\ \rm{Mpc}^{-1}$ and $300\ \rm{km\ s}^{-1}\ \rm{Mpc}^{-1}$.
We have already mentioned that we need to assume some uncertainty in the EoS parameters.
The projected uncertainty in $R_{1.4}$ measurement is $\sim 1$ km after the fifth observing (O5) run of LIGO-Virgo-KAGRA network~\cite{Landry:2020vaw}. 
The constraint on the NS radius is expected to be better than $0.1$ km from  BNS observations  by the time of CE \cite{Evans:2021gyd}. 
So, we consider three sets of uncertainties in the EoS parameters, mentioned in Table~\ref{tab:EoS_prior} along with the injected values of the EoS parameters.
The mass-radius plot for all the prior EoSs, shown in Fig.~\ref{fig:eos_prior} gives a similar qualitative measurement about the uncertainties in the EoS parameters we are considering in our work.
Now, the posterior of Hubble constant along with the EoS parameters have been measured using the redshift information and luminosity distance of GW sources using the Bayesian framework as described above.

\begin{figure*}
    \centering
    \includegraphics [width = 17cm,height=6cm] {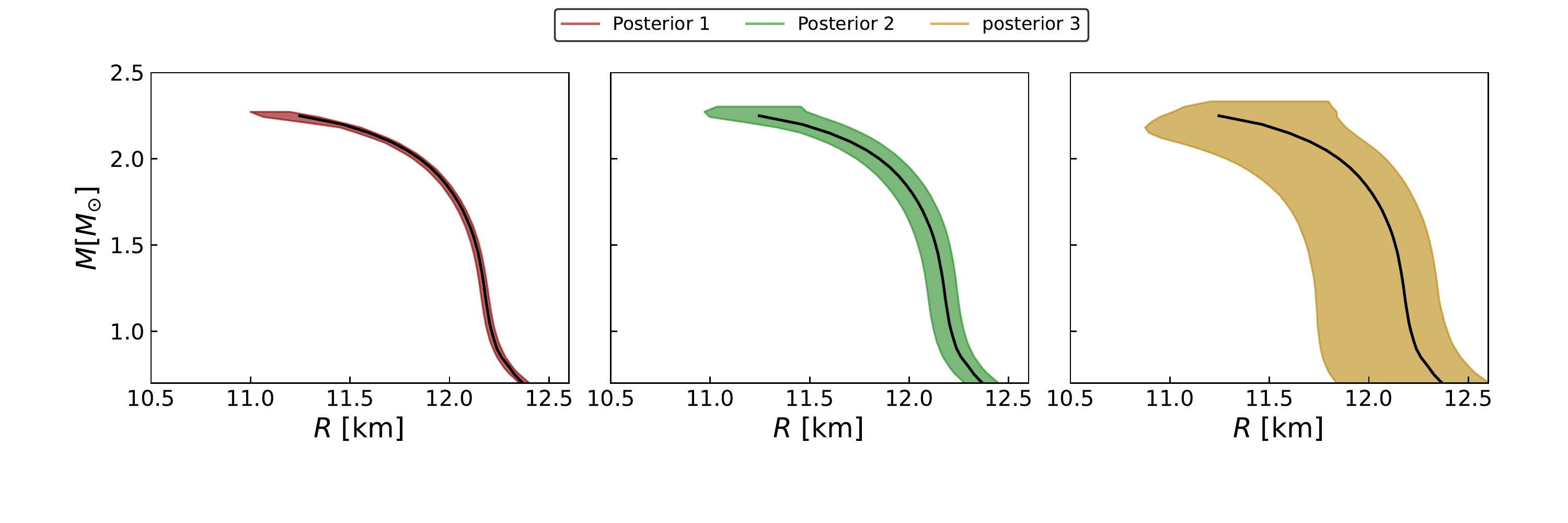}
    \caption{$90\%$ credible interval of mass-radius posteriors are shown using three different choices of the EoS prior distributions as mentioned in Table \ref{tab:EoS_prior}. The solid black curve in each plot is the mass-radius plot corresponding to the injected EoS parameters.}
    \label{fig:eos_post}
\end{figure*}

\begingroup
\renewcommand{\arraystretch}{1.4}
\begin{table*}
\caption{\label{tab:EoS_post}
Median and $90\%$ credible interval of the EoS Parameters, $R_{1.4}$, $R_{2}$, $\Lambda_{1.4}$, $\Lambda_{2}$, $M_{\text{max}}$, skin thickness and Hubble constant as obtained using our methodology to $50$ GW events with EoS Prior 1, Prior 2 and Prior 3, mentioned in Table~\ref{tab:EoS_prior}.
The parameters, such as $R_{1.4}$, $R_{2}$, $M_{\text{max}}$ and skin thickness have been inferred from the injected EoS parameters ($K_{0}, e_{sym}, L, K_{sym}, \Gamma_{1}, \Gamma_{2}, \Gamma_{1}$).}
\begin{ruledtabular}
\begin{tabular} {ccccc}
EoS Parameters & Injected & Posterior 1 & Posterior 2 & Posterior 3 \\
\colrule
$K_{0}$ [MeV] & $240$ & $240.00^{+0.69}_{-0.69}$ & $240.16^{+5.30}_{-5.51}$ & $239.17^{+7.52}_{-7.45}$ \\
$e_{sym}$ [MeV] & $31.70$ & $31.67^{+0.13}_{-0.13}$ & $31.74^{+0.68}_{-0.69}$ & $31.68^{+0.82}_{-0.82}$ \\
$L$ [MeV] & $58.70$ & $58.63^{+0.40}_{-0.39}$ & $58.56^{+1.33}_{-1.39}$ & $56.47^{+6.77}_{-6.76}$ \\
$K_{sym}$ [MeV] & $-100$ & $-100.02^{+0.67}_{-0.69}$ & $-100.99^{+13.65}_{-13.65}$ & $-109.15^{+23.79}_{-24.19}$ \\
$\Gamma_{1}$ & $2.5$ & $2.50^{+0.01}_{-0.01}$ & $2.49^{+0.06}_{-0.06}$ & $2.47^{+0.12}_{-0.13}$ \\
$\Gamma_{2}$ & $3.5$ & $3.50^{+0.01}_{-0.01}$ & $3.51^{+0.06}_{-0.06}$ & $3.51^{+0.14}_{-0.14}$ \\
$\Gamma_{3}$ & $3.0$ & $3.00^{+0.13}_{-0.13}$ & $3.19^{+0.34}_{-0.36}$ & $3.11^{+0.36}_{-0.47}$ \\
$R_{1.4 M_{\odot}}$ [km] & $12.153$ & $12.151^{+0.019}_{-0.019}$ & $12.145^{+0.077}_{-0.077}$ & $12.029^{+0.297}_{-0.321}$ \\ 
$R_{2 M_{\odot}}$ [km] & $11.851$ & $11.847^{+0.026}_{-0.027}$ & $11.851^{+0.107}_{-0.114}$ & $11.713^{+0.372}_{-0.439}$ \\
$\Lambda_{1.4 M_{\odot}}$ & $375.813$ & $379.829^{+6.469}_{-6.266}$ & $378.997^{+17.064}_{-16.702}$ & $357.726^{+55.797}_{-51.555}$ \\ 
$\Lambda_{2 M_{\odot}}$ & $29.33$ & $29.243^{+0.570}_{0.585}$ & $29.419^{+2.392}_{-2.404}$ & $26.901^{+7.506}_{-7.335}$ \\ 
$M_{\text{max}}$ [$M_{\odot}$] & $2.271$ & $2.267^{+0.008}_{-0.011}$ & $2.247^{+0.041}_{-0.069}$ & $2.221^{+0.092}_{-0.092}$ \\ 
$R_{\text{skin}}^{208}$ [fm] & $0.1873$ & $0.1872^{+0.0006}_{-0.0006}$ & $0.1871^{+0.0019}_{-0.0020}$ & $0.1840^{+0.0099}_{-0.0099}$ \\
$R_{\text{skin}}^{48}$ [fm] & $0.1571$ & $0.1571^{+0.0004}_{-0.0003}$ & $0.1571^{+0.0012}_{-0.0012}$ & $0.1551^{+0.0061}_{-0.0061}$ \\
$H_{0}$ [km/s/Mpc] & $70$ & $70.21^{+5.37}_{-4.95}$ & $69.30^{+5.42}_{-5.35}$ & $72.96^{+8.10}_{-7.84}$ \\
\end{tabular}
\end{ruledtabular}
\end{table*}
\endgroup

\begin{figure}
    \centering
    \includegraphics[width = 8.6cm,height=6cm] {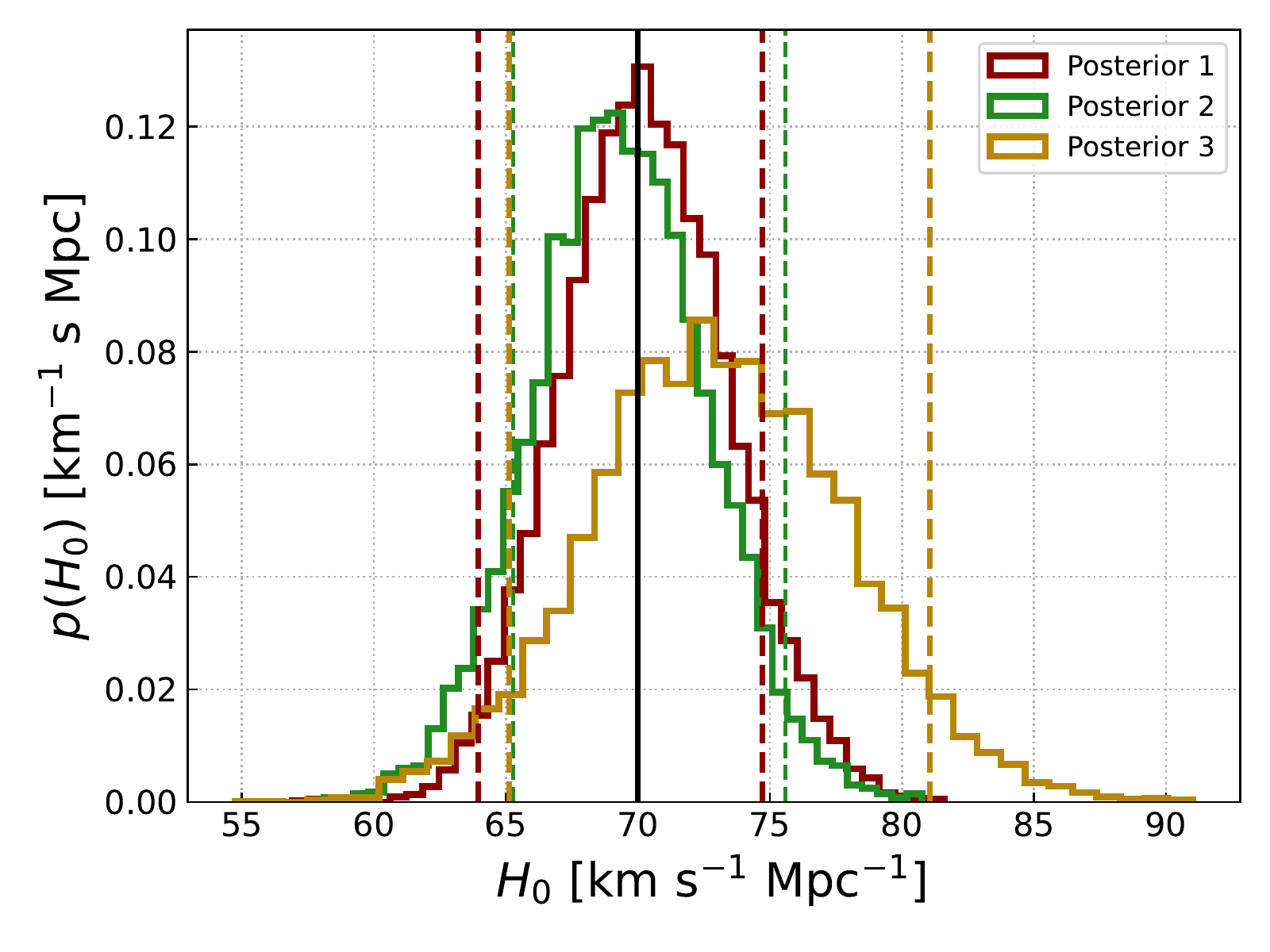}
    \caption{Comparison of joint estimate of the Hubble constant over $50$ GW events for the different choices of EoS. The black vertical line denotes the injected value of $H_{0}=70\ \rm{km\ s}^{-1}\ \rm{Mpc}^{-1}$. The dashed vertical lines of the same color indicates $90\%$ highest density credible interval corresponding to the histogram of that color. }
    \label{fig:H0_comb}
\end{figure}

\section{Results} \label{results}
\begin{figure}
    \centering
    \includegraphics [width = 8.6cm,height=18cm] {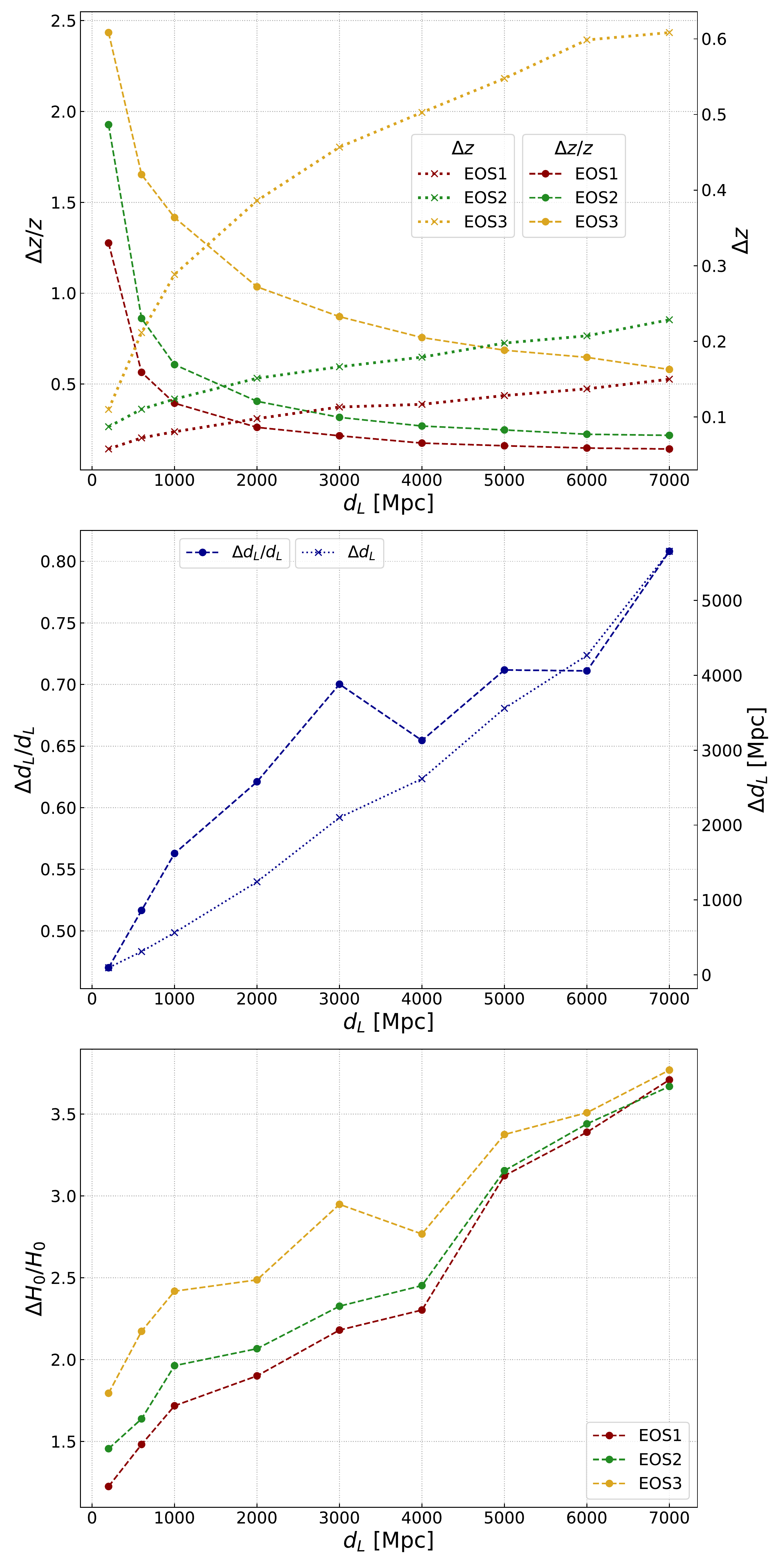}
    \caption{Comparison of uncertainty in the measurement of $z$, $d_L$, and $H_0$ using a single BNS observation with source-frame mass pair $(1.4+1.26) M_{\odot}$ at different representative distances. \emph{Top Panel}: fractional uncertainty (left vertical axis) and uncertainty (right vertical axis) in the redshift measurement for three choices of EoS prior. \emph{Middle Panel}: fractional uncertainty (left vertical axis) and uncertainty (right vertical axis) in estimating luminosity distance from GW data. \emph{Bottom Panel}: fractional uncertainty of the Hubble constant for different EoS priors considering uniform prior between $10\ \rm{km\ s}^{-1}\ \rm{Mpc}^{-1}$ and $400\ \rm{km\ s}^{-1}\ \rm{Mpc}^{-1}$}.
    \label{fig:error_compariosn}
\end{figure}

\begin{figure}
    \centering
    \includegraphics [width = 8.6cm,height=6cm] {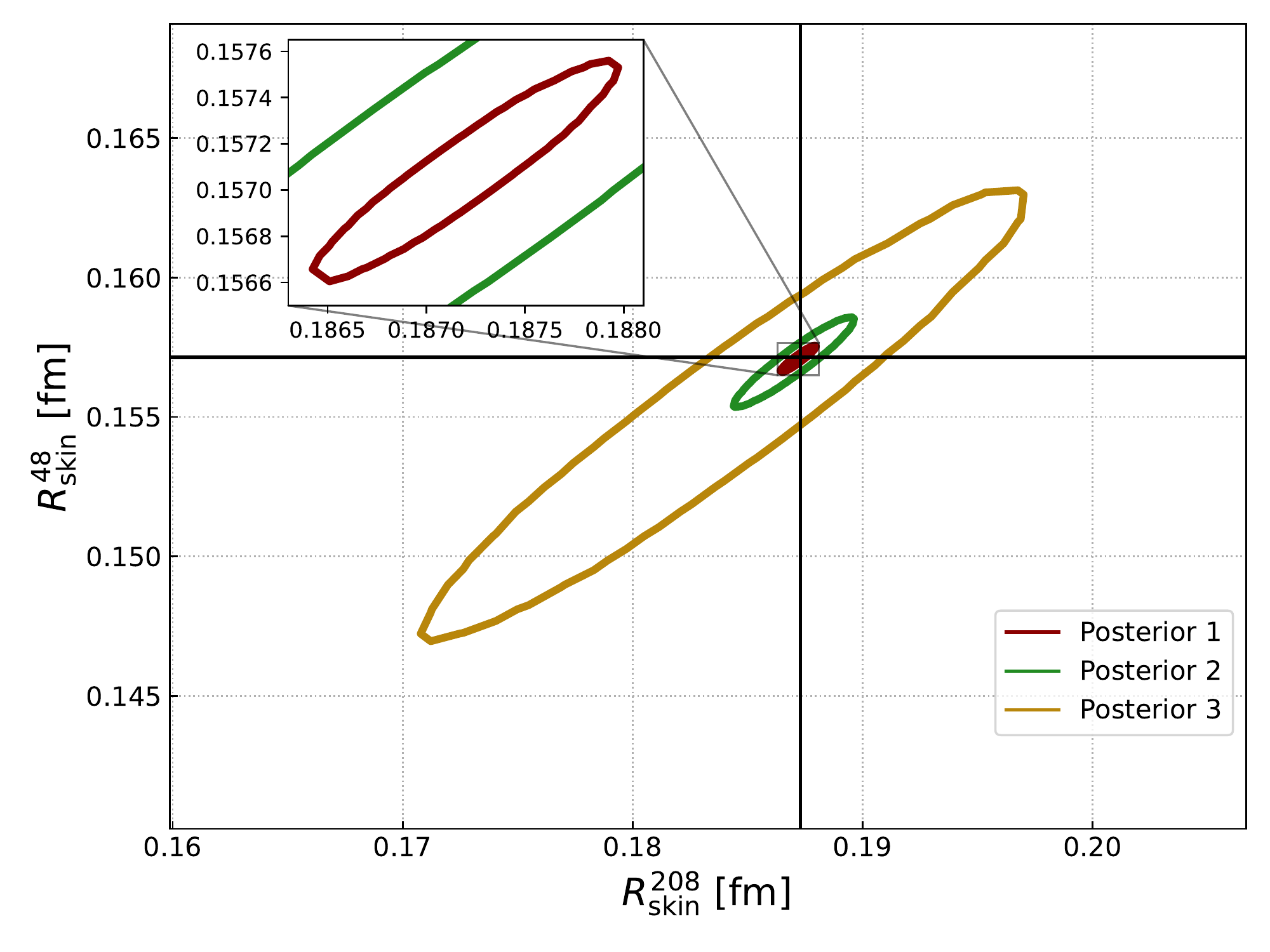}
    \caption{Inferred joint posterior distributions ($90\%$ credible interval) of $R_{\text{skin}}^{48}$ and $R_{\text{skin}}^{208}$ is shown for the different choices of the EoS priors. The black vertical and horizontal lines denote the injected value of the skin thicknesses $R_{\text{skin}}^{208}= 0.1873$ fm and $R_{\text{skin}}^{48} = 0.1571 $ fm respectively, corresponding to injected EoS.}
    \label{fig:skin}
\end{figure}

We apply the method described in Sec.~\ref{method} to $50$ GW events to test its efficacy 
in simultaneously estimating the EoS parameters and the Hubble constant. We consider a simulated GW signal as detected if its network SNR turns out to be equal to or greater than $8$.
The individual posteriors of the EoS parameters and the Hubble constant may not always be peaked at the true values; some posteriors may be multimodal as well.
In particular, there exists a degeneracy in the measurability of the luminosity distance and the inclination angle of binaries in GWs~\cite{Cutler:1994ys, LIGOScientific:2013yzb}; e.g., a source will appear fainter not just when it is farther but also when its orbital plane is more tilted relative to the line of sight (see Ref.~\cite{Chen:2017rfc}).
Since the orbit orientations in a population are random, the expectation is that with an increasing number of observations, this method of combining individual source parameter posteriors will still succeed in constraining $H_{0}$ and the EoS parameters with increasing precision since every BNS signal is characterized by the same $H_{0}$ and the EoS parameters.

The combined posteriors of the inferred EoS parameters and $H_{0}$, together with their correlation, are shown in Figs.~\ref{fig:CE_EOS1},~\ref{fig:CE_EOS2} and~\ref{fig:CE_EOS3} corresponding to Priors $1$, $2$ and $3$, respectively.
The different choices of priors in the EoS parameters (also discussed in Sec.~\ref{simulation}) are mentioned in Table~\ref{tab:EoS_prior}.
We summarize the $90\%$ credible interval of the inferred EoS parameters and Hubble constant in Table~\ref{tab:EoS_post}. 
We also mention some other inferred parameters from the NS EoS posteriors like $R_{1.4M_{\odot}}$ (radius of neutron star at $1.4 M_{\odot}$), $R_{2M_{\odot}}$, $\Lambda_{1.4 M_{\odot}}$ (tidal deformability of neutron star at $1.4 M_{\odot}$) and $\Lambda_{2M_{\odot}}$ in Table~\ref{tab:EoS_post}.

From the joint posterior of $H_{0}$ and the EoS parameters (see Fig.~\ref{fig:CE_EOS1},~\ref{fig:CE_EOS2} and~\ref{fig:CE_EOS3} and Table~\ref{tab:EoS_post}), it is 
evident that the uncertainty in the measurement of the Hubble constants using EoS Priors $1$ and $2$ are not significantly different (see Fig.~\ref{fig:H0_comb}).
In comparison, the posterior of the Hubble constant estimated using EoS Prior $3$ is slightly broad.
However, the uncertainties in the posterior of the EoS parameters are quite different: To wit, a tighter prior of the EoS parameters leads to narrower EoS parameter posteriors.
The mass-radius plots (see Fig.~\ref{fig:eos_post} and Table~\ref{tab:EoS_post} for the median values with the $90\%$ credible intervals of EoS parameters) corresponding to three EoS posteriors also reveal that the constraints of the EoS parameters are significantly different. 

We have also studied the (fractional) uncertainty in inferring the redshift and the Hubble constant with BNSs (with source-frame masses $(1.4+1.26) M_{\odot}$) as functions of the luminosity distances. The results are shown in Fig.~\ref{fig:error_compariosn}. 
How precisely one is able to measure the redshift depends on the choice of EoS prior (see the top panel of Fig.~\ref{fig:error_compariosn}).
The errors in the measured distances of GW events are also expected to increase with the distances of the sources, as shown in the bottom panel of Fig.~\ref{fig:error_compariosn}.
In comparison, the large errors in the inferred distances of far away sources, as opposed to the nearer sources, dominate the errors in their redshift estimates. 
Consequently, the uncertainty in the Hubble constant for GW events from large distances is not highly reliant on the uncertainties in EoS parameters (see the bottom panel of Fig.~\ref{fig:error_compariosn}).
So, the combined estimation of the Hubble constant in our study, considering all the events, is weakly dependent on the EoS parameters' uncertainty, as the BNS distribution (uniform in comoving volume and source-frame time) in this work is dominated by distant sources.
We can also expect some improvement in the mitigation of the well-known distance-inclination degeneracy with the development of the higher-mode waveform models for BNSs~\cite{CalderonBustillo:2020kcg}, thereby aiding a more precise estimation of the Hubble constant.

Chatterjee \textit{et al.}~\cite{Chatterjee:2021xrm} mentioned that $\Delta H_{0}/H_{0} \sim 2\%$ for $N_{\text{det}} = 10^{3}$ detected GW events observed by CE. 
They assumed $\sim 1/\sqrt{N_{\text{det}}}$ scaling for estimating the projected uncertainty measurement in the Hubble constant during CE era. 
We find the $90\%$ credible interval error $\Delta H_{0}/H_{0} \lesssim 15-22\%$ (see Fig.~\ref{fig:H0_comb}), depending on the prior EoS, while inferring the Hubble constant from $50$ BNS observations.
We also estimate the uncertainty with a similar scaling $\sim 1/\sqrt{N_{\text{det}}}$ in the measurement of the Hubble constant $\Delta H_{0}/H_{0} \lesssim 3-5\%$ with $N_{\text{det}}=10^{3}$ events, depending on the uncertainties in the EoS parameters.
The projected uncertainty in $H_{0}$ measurement is slightly greater than the uncertainty estimation of $H_{0}$, mentioned by Chatterjee \textit{et al.}~\cite{Chatterjee:2021xrm}. 
It is expected because we consider uncertainties in the EoS parameters, unlike Chatterjee \textit{et al.}~\cite{Chatterjee:2021xrm}.
However, improvement in the uncertainty due to luminosity distance-inclination angle degeneracy should have a more significant effect in constraining $H_{0}$ as discussed above.

Recently, PREX-II collaboration has reported~\cite{PREX:2021umo} the value of neutron skin thickness of $^{208} {\rm Pb}$ to be, $R_{\rm skin}^{208} = 0.29 \pm 0.07$ fm (mean and $1 \sigma$ standard deviation). Ref.~\cite{Reed:2021nqk} have argued, this result implies the value of $L$ to be $106 \pm 37$ MeV. They also deduce a larger value of NS radius as there exists a correlation between $R_{1.4}$ and $L$ based on relativistic mean field calculation. However, Ref.~\cite{Biswas:2021yge} argued that the measurement uncertainty in $R_{\rm skin}^{208}$ by PREX-II is quite broad; Moreover, the measurement uncertainty deduced after the addition of astrophysical observations is dominated by the latter.
Combined astrophysical observations and PREX-II data yield the value of empirical parameter $L = 69^{+21}_{-19}$ MeV, $R_{\rm skin}^{208} = 0.20_{-0.05}^{+0.05}$ fm, and radius of a $1.4 M_{\odot}$ ($R_{1.4}) = 12.70_{-0.54}^{+ 0.42}$ km  at $1 \sigma$ credible interval. Nevertheless, a better measurement of $R_{\rm skin}^{208}$ might have a small effect on the radius of low mass NSs, but for the high masses, there will be almost no effect. The state-of-the-art chiral effective field theory calculations predict~\cite{Xu:2020fdc} the value of $R_{\rm skin}$ to be $0.17-0.18$ fm based on Bayesian analyses using mocked data. Therefore, if the high $R_{\rm skin}^{208}$ values continue to persist with lesser uncertainty, it would pose a challenge to the current theoretical understanding about the nuclear matter near the saturation densities. More recently, the CREX collaboration~\cite{CREX:2022kgg} has reported the value of neutron skin thickness of $\rm ^{48} \! Ca$ to be $R_{\rm skin}^{48} = 0.121 \pm 0.026$ (exp) $\pm 0.024$ (model) fm. They find  several models, including the microscopic
coupled cluster calculations~\cite{Hu:2021trw} are consistent with the combined CREX and PREX-II results at $90 \%$ credible interval, but in tension at $68 \%$ credible interval.
Following Ref.~\cite{Biswas:2021yge} we estimate the skin thickness $R_{\text{skin}}$ using the universal relation from Vi\~nas \textit{et al}.~\cite{Vinas:2013hua} connecting $R_{\text{skin}}$ and the empirical parameter $L$: $R_{\text{skin}}^{208} [\text{fm}] = 0.101 +  0.00147 \times L [\text{MeV}]$. 
Then we also use another empirical relation from Ref.~\cite{Tripathy:2020yig}, $R_{\text{skin}}^{48} =  0.0416 +  0.6169 R_{\text{skin}}^{208}$, to obtain the neutron skin thickness of $^{48} \text{Ca}$.
The estimation of both skin thicknesses for the different EoS priors are shown in Fig.~\ref{fig:skin} and Table~\ref{tab:EoS_post}. We find irrespective of the choice in the EoS priors, future observations from CE will constrain $R_{\rm skin}$ with subpercentage precision. So if indeed there is a tension between nuclear theory and astrophysical observations, it will be revealed.

\section{Discussion} \label{discussion}

In this paper, we demonstrate how a population of BNS can constrain the EoS parameters and the Hubble constant simultaneously during 3rd generation detector era
when unique transient electromagnetic counterparts cannot be identified.
For dark GW signal from CBC, the current state of the art is to perform a statistical approach by considering the redshifts of all potential host galaxies within the three-dimensional localization region of GW event to get a constraint on the distance-redshift relationship~\cite{Schutz:1986gp, Chen:2017rfc, LIGOScientific:2018gmd, Gray:2019ksv}.
However, our method can infer the redshift information of the GW event from the knowledge of the prior EoS parameters. We combine the redshift information with the luminosity distance measured from GW observation to estimate the Hubble constant.

We have tested our formalism using three different priors of the EoS parameters. 
We only infer the Hubble constant as the cosmological parameter in our Bayesian analysis. 
Since the $3$rd generation detectors will have better sensitivity and observe many more 
sources, it is conceivable that they will be able to constrain additional cosmological parameters apart from $H_{0}$.
However, one drawback of this work is that we took the  merger rate to be constant and employed a simple mass population model of BNS distribution.
Ideally, one should use a more realistic merger rate (that evolves with redshift) as well as a mass population model of BNS distribution~\cite{Alsing:2017bbc, Landry:2021hvl}.
The merger rate, along with horizon redshift, determines the total number of BNS mergers, which subsequently gives an estimate of the uncertainty in EoS parameters and the Hubble constant.
Since the projected horizon redshift $z \sim 11$ for CE~\cite{Evans:2021gyd} is quite large, the total number of mergers is sensitive to the choice of merger rate. 
In addition to the merger rate, the mass-population model also determines the SNR distribution of GW events. Consequently, the number of detected BNS events also depends on the mass population.
So, it is important to consider a realistic merger rate and mass population model of BNS to predict the measurement uncertainty of EoS parameters and $H_{0}$
that can be achieved during the $3$rd generation detector era.

Our methodology can easily be extended to incorporate simultaneous inference of cosmological parameters, population model parameters and merger rate parameters of NSs.
It is also important to note that in our formalism, the ability to constrain $H_{0}$ and NS EoS using BNS signals reduces with increasing GW source distance. 
This is because the errors in certain source parameters, especially luminosity distance and tidal deformability, worsen for distant sources.
As one observes farther sources with 3G instruments, their numbers in a spherical shell of constant comoving thickness will increase, at least up to some redshift~\cite{LIGOScientific:2021psn}.
Thus, even if the error in $H_0$ from individual sources worsens with increasing  distance, as shown in Fig.~\ref{fig:error_compariosn}, nevertheless, the error in the combined estimate of $H_0$ from all sources within the shell would worsen less.
The role of population models and the merger rate on $H_0$ measurements will be studied in more detail in a separate work~\cite{dark_bns}.

In the case of the EoS parameters, other observations can also put stringent measurements of the mass and radii of NSs.
In particular, the observation of PSR J0030+0451~\cite{Riley:2019yda,Miller:2019cac} PSR J0740+6620~\cite{Riley:2021pdl,Miller:2021qha} by NICER collaboration has been utilized to measure mass-radius relation of the NS.

\section*{Acknowledgments}

We thank Rana Nandi and Prasanta Char for earlier collaboration (with B.B. and S.B.) on the construction of the neutron star equation of state parametrization used here and for useful discussions.
We would also like to thank Hsin-Yu Chen for carefully reading the manuscript and making several useful suggestions.
T.G. gratefully acknowledges using the LDG clusters, Sarathi at IUCAA and Hawk at Cardiff University, accessed through the LIGO-Virgo-Kagra Collaboration.
We are also grateful for the computational resource Pegasus provided by IUCAA for this work. 
B.B. also acknowledges the support from the Knut and Alice Wallenberg Foundation 
under grant Dnr. KAW 2019.0112.

\appendix

\section{Bayesian Framework} \label{bayesian_details}

From Bayes' theorem, the joint posterior of cosmological parameters and the EoS parameters for a detected GW event $p(H_{0}, \bm{\mathcal{E}} \mid  x_{\rm{GW}} )$ is given by
\begin{equation} \label{eq1}
    p(H_{0}, \bm{\mathcal{E}} \mid  x_{\rm{GW}} ) \propto p(x_{\rm{GW}} \mid  H_{0}, \bm{\mathcal{E}} ) \pi(H_{0}) \pi(\bm{\mathcal{E}})\,.
\end{equation}

The likelihood for a single GW event in Eq.~\eqref{eq1} can be written as 
\begin{widetext} 
\begin{eqnarray} \label{likelihood1}
    p (x_{\rm{GW}}\mid  H_{0}, \bm{\mathcal{E}}) \propto \frac{1}{\beta (H_{0})} && \iiint p \big(x_{\rm GW} \mid m_{1}^{z}(m_{1},z), m_{2}^{z}(m_{2},z), \Lambda_{1}(m_{1}, \bm{\mathcal{E}}), \Lambda_{2}(m_{2}, \bm{\mathcal{E}}), d_{L} (z, H_{0}) \big)\nonumber \\
    && \times \pi(m_{1}, m_{2} \mid \bm{\mathcal{E}}) \pi(z)  dz dm_{1} dm_{2} \,.
\end{eqnarray}
\end{widetext}
In terms of explicit parameters, Eq.~\eqref{likelihood1} can conveniently be written as 
\begin{eqnarray}
    && p(x_{\rm{GW}} \mid  H_{0}, \bm{\mathcal{E}} ) \nonumber \\ && \propto \frac{1}{\beta(H_{0})} \iiint  p(x_{\rm GW} \mid z,m_{1}, m_{2}, \bm{\mathcal{E}}, H_{0}) \nonumber \\
    && \times \pi(m_{1}, m_{2} \mid \bm{\mathcal{E}}) \pi(z)  dz dm_{1} dm_{2}\,,
\end{eqnarray}
where a normalization term $\beta(H_{0})$ has been included in the denominator to account for selection effects~\cite{Mandel:2018mve, Chen:2017rfc}.

\begin{equation} \label{beta2}
    \beta(H_{0}) = \int p_{\rm{det}} (z, H_{0}) \pi (z) dz \,.
\end{equation}
The term $p_{\rm{det}} (z, H_{0}) $ in Eq.~\eqref{beta2} denotes detection probability for a particular choice of redshift and $H_{0}$. 
We elaborate the calculation of $p_{\rm{det}} (z, H_{0})$ (similar to $p(d_{\rm{GW}} \mid  z, H_{0})$ in Ref.~\cite{Gray:2019ksv}) following Gray \textit{et al.}~\cite{Gray:2019ksv}
(implemented in \verb+gwcosmo+~\footnote{\label{gwcosmo}gwcosmo code: \url{https://git.ligo.org/lscsoft/gwcosmo}}).

The final joint posterior of the Hubble constant and the EoS parameters can be obtained by combining sources, as follows:
\begin{equation} \label{joint_pH0}
    p(H_{0}, \bm{\mathcal{E}} \mid  \{x_{\rm{GW}_{i}}\} ) = \prod_{i=1}^{N_{\rm{det}}} p(H_{0}, \bm{\mathcal{E}} \mid  x_{\rm{GW}_i} ) \,.
\end{equation}

\section{Estimation of Detection Probability} \label{selection_effect}

The selection effect as defined in Eq.~\eqref{beta} [also in Eq.~\eqref{beta2}] completely depends on GW detection efficiency. GW detector can identify those signals that generate sufficiently high amplitude response. We consider a GW event as detected if the network SNR is some threshold SNR $\rho_{\text{th}}$ or higher.
In our work, we set $\rho_{\text{th}}=8$.
We calculate $p_{\text{det}}$ using the injection of BNS signal.
We populate $10^{3}$ BNSs uniformly over the entire sky for each of the redshift bin and a given choice of Hubble constant.
These BNSs follow the same population model as we assumed for this work.
We have injected the total number of BNS mergers $\sim 10^{7}$ for the entire range of the Hubble constant and redshift for computing detection probability; among all the sources, $\sim 8 \times 10^{6}$ sources qualify the detection criterion.
We then calculate the matched filtering SNR~\cite{Sathyaprakash:2009xs} of the injected GW signal of $128$s at each of the detector and hence the network SNR. For this purpose, we have used \verb+bilby+~\cite{Ashton_2019} to inject GW signal and calculate the SNR.
Since, $\rho^{2}$ follows noncentral chi-squared distribution with two degrees of freedom~\cite{LIGOScientific:2019hgc}, in the presence of stationary and Gaussian noise with a known power spectrum, we thus estimate the detection probability $p_{\text{det}}$ for a certain redshift bin and a particular choice of $H_{0}$ by probability density function of noncentral chi-squared distribution with degrees of freedom is twice of the number of detectors and noncentrality parameter is $\rho_{th}$. The detection probability is estimated in the similar way as implemented in \verb+gwcosmo+.
Now, it is straight forward to calculate the selection function $\beta(H_{0})$ [Eq.~\eqref{beta}] by marginalizing the detection probability over the redshift prior used in the Bayesian framework.

\bibliography{citations}

\begin{thebibliography}{70}%
\makeatletter
\providecommand \@ifxundefined [1]{%
 \@ifx{#1\undefined}
}%
\providecommand \@ifnum [1]{%
 \ifnum #1\expandafter \@firstoftwo
 \else \expandafter \@secondoftwo
 \fi
}%
\providecommand \@ifx [1]{%
 \ifx #1\expandafter \@firstoftwo
 \else \expandafter \@secondoftwo
 \fi
}%
\providecommand \natexlab [1]{#1}%
\providecommand \enquote  [1]{``#1''}%
\providecommand \bibnamefont  [1]{#1}%
\providecommand \bibfnamefont [1]{#1}%
\providecommand \citenamefont [1]{#1}%
\providecommand \href@noop [0]{\@secondoftwo}%
\providecommand \href [0]{\begingroup \@sanitize@url \@href}%
\providecommand \@href[1]{\@@startlink{#1}\@@href}%
\providecommand \@@href[1]{\endgroup#1\@@endlink}%
\providecommand \@sanitize@url [0]{\catcode `\\12\catcode `\$12\catcode
  `\&12\catcode `\#12\catcode `\^12\catcode `\_12\catcode `\%12\relax}%
\providecommand \@@startlink[1]{}%
\providecommand \@@endlink[0]{}%
\providecommand \url  [0]{\begingroup\@sanitize@url \@url }%
\providecommand \@url [1]{\endgroup\@href {#1}{\urlprefix }}%
\providecommand \urlprefix  [0]{URL }%
\providecommand \Eprint [0]{\href }%
\providecommand \doibase [0]{http://dx.doi.org/}%
\providecommand \selectlanguage [0]{\@gobble}%
\providecommand \bibinfo  [0]{\@secondoftwo}%
\providecommand \bibfield  [0]{\@secondoftwo}%
\providecommand \translation [1]{[#1]}%
\providecommand \BibitemOpen [0]{}%
\providecommand \bibitemStop [0]{}%
\providecommand \bibitemNoStop [0]{.\EOS\space}%
\providecommand \EOS [0]{\spacefactor3000\relax}%
\providecommand \BibitemShut  [1]{\csname bibitem#1\endcsname}%
\let\auto@bib@innerbib\@empty
\bibitem [{\citenamefont {Schutz}(1986)}]{Schutz:1986gp}%
  \BibitemOpen
  \bibfield  {author} {\bibinfo {author} {\bibfnamefont {B.~F.}\ \bibnamefont
  {Schutz}},\ }\href {\doibase 10.1038/323310a0} {\bibfield  {journal}
  {\bibinfo  {journal} {Nature}\ }\textbf {\bibinfo {volume} {323}},\ \bibinfo
  {pages} {310} (\bibinfo {year} {1986})}\BibitemShut {NoStop}%
\bibitem [{\citenamefont {Chen}\ \emph {et~al.}(2018)\citenamefont {Chen},
  \citenamefont {Fishbach},\ and\ \citenamefont {Holz}}]{Chen:2017rfc}%
  \BibitemOpen
  \bibfield  {author} {\bibinfo {author} {\bibfnamefont {H.-Y.}\ \bibnamefont
  {Chen}}, \bibinfo {author} {\bibfnamefont {M.}~\bibnamefont {Fishbach}}, \
  and\ \bibinfo {author} {\bibfnamefont {D.~E.}\ \bibnamefont {Holz}},\ }\href
  {\doibase 10.1038/s41586-018-0606-0} {\bibfield  {journal} {\bibinfo
  {journal} {Nature}\ }\textbf {\bibinfo {volume} {562}},\ \bibinfo {pages}
  {545} (\bibinfo {year} {2018})},\ \Eprint {http://arxiv.org/abs/1712.06531}
  {arXiv:1712.06531 [astro-ph.CO]} \BibitemShut {NoStop}%
\bibitem [{\citenamefont {Fishbach}\ \emph {et~al.}(2019)\citenamefont
  {Fishbach} \emph {et~al.}}]{LIGOScientific:2018gmd}%
  \BibitemOpen
  \bibfield  {author} {\bibinfo {author} {\bibfnamefont {M.}~\bibnamefont
  {Fishbach}} \emph {et~al.} (\bibinfo {collaboration} {LIGO Scientific,
  Virgo}),\ }\href {\doibase 10.3847/2041-8213/aaf96e} {\bibfield  {journal}
  {\bibinfo  {journal} {Astrophys. J. Lett.}\ }\textbf {\bibinfo {volume}
  {871}},\ \bibinfo {pages} {L13} (\bibinfo {year} {2019})},\ \Eprint
  {http://arxiv.org/abs/1807.05667} {arXiv:1807.05667 [astro-ph.CO]}
  \BibitemShut {NoStop}%
\bibitem [{\citenamefont {Nair}\ \emph {et~al.}(2018)\citenamefont {Nair},
  \citenamefont {Bose},\ and\ \citenamefont {Saini}}]{Nair:2018ign}%
  \BibitemOpen
  \bibfield  {author} {\bibinfo {author} {\bibfnamefont {R.}~\bibnamefont
  {Nair}}, \bibinfo {author} {\bibfnamefont {S.}~\bibnamefont {Bose}}, \ and\
  \bibinfo {author} {\bibfnamefont {T.~D.}\ \bibnamefont {Saini}},\ }\href
  {\doibase 10.1103/PhysRevD.98.023502} {\bibfield  {journal} {\bibinfo
  {journal} {Phys. Rev. D}\ }\textbf {\bibinfo {volume} {98}},\ \bibinfo
  {pages} {023502} (\bibinfo {year} {2018})},\ \Eprint
  {http://arxiv.org/abs/1804.06085} {arXiv:1804.06085 [astro-ph.CO]}
  \BibitemShut {NoStop}%
\bibitem [{\citenamefont {Gray}\ \emph {et~al.}(2020)\citenamefont {Gray} \emph
  {et~al.}}]{Gray:2019ksv}%
  \BibitemOpen
  \bibfield  {author} {\bibinfo {author} {\bibfnamefont {R.}~\bibnamefont
  {Gray}} \emph {et~al.},\ }\href {\doibase 10.1103/PhysRevD.101.122001}
  {\bibfield  {journal} {\bibinfo  {journal} {Phys. Rev. D}\ }\textbf {\bibinfo
  {volume} {101}},\ \bibinfo {pages} {122001} (\bibinfo {year} {2020})},\
  \Eprint {http://arxiv.org/abs/1908.06050} {arXiv:1908.06050 [gr-qc]}
  \BibitemShut {NoStop}%
\bibitem [{\citenamefont {Borhanian}\ \emph {et~al.}(2020)\citenamefont
  {Borhanian}, \citenamefont {Dhani}, \citenamefont {Gupta}, \citenamefont
  {Arun},\ and\ \citenamefont {Sathyaprakash}}]{Borhanian:2020vyr}%
  \BibitemOpen
  \bibfield  {author} {\bibinfo {author} {\bibfnamefont {S.}~\bibnamefont
  {Borhanian}}, \bibinfo {author} {\bibfnamefont {A.}~\bibnamefont {Dhani}},
  \bibinfo {author} {\bibfnamefont {A.}~\bibnamefont {Gupta}}, \bibinfo
  {author} {\bibfnamefont {K.~G.}\ \bibnamefont {Arun}}, \ and\ \bibinfo
  {author} {\bibfnamefont {B.~S.}\ \bibnamefont {Sathyaprakash}},\ }\href
  {\doibase 10.3847/2041-8213/abcaf5} {\bibfield  {journal} {\bibinfo
  {journal} {Astrophys. J. Lett.}\ }\textbf {\bibinfo {volume} {905}},\
  \bibinfo {pages} {L28} (\bibinfo {year} {2020})},\ \Eprint
  {http://arxiv.org/abs/2007.02883} {arXiv:2007.02883 [astro-ph.CO]}
  \BibitemShut {NoStop}%
\bibitem [{\citenamefont {Bera}\ \emph {et~al.}(2020)\citenamefont {Bera},
  \citenamefont {Rana}, \citenamefont {More},\ and\ \citenamefont
  {Bose}}]{Bera:2020jhx}%
  \BibitemOpen
  \bibfield  {author} {\bibinfo {author} {\bibfnamefont {S.}~\bibnamefont
  {Bera}}, \bibinfo {author} {\bibfnamefont {D.}~\bibnamefont {Rana}}, \bibinfo
  {author} {\bibfnamefont {S.}~\bibnamefont {More}}, \ and\ \bibinfo {author}
  {\bibfnamefont {S.}~\bibnamefont {Bose}},\ }\href {\doibase
  10.3847/1538-4357/abb4e0} {\bibfield  {journal} {\bibinfo  {journal}
  {Astrophys. J.}\ }\textbf {\bibinfo {volume} {902}},\ \bibinfo {pages} {79}
  (\bibinfo {year} {2020})},\ \Eprint {http://arxiv.org/abs/2007.04271}
  {arXiv:2007.04271 [astro-ph.CO]} \BibitemShut {NoStop}%
\bibitem [{\citenamefont {Mukherjee}\ \emph {et~al.}(2021)\citenamefont
  {Mukherjee}, \citenamefont {Wandelt}, \citenamefont {Nissanke},\ and\
  \citenamefont {Silvestri}}]{Mukherjee:2020hyn}%
  \BibitemOpen
  \bibfield  {author} {\bibinfo {author} {\bibfnamefont {S.}~\bibnamefont
  {Mukherjee}}, \bibinfo {author} {\bibfnamefont {B.~D.}\ \bibnamefont
  {Wandelt}}, \bibinfo {author} {\bibfnamefont {S.~M.}\ \bibnamefont
  {Nissanke}}, \ and\ \bibinfo {author} {\bibfnamefont {A.}~\bibnamefont
  {Silvestri}},\ }\href {\doibase 10.1103/PhysRevD.103.043520} {\bibfield
  {journal} {\bibinfo  {journal} {Phys. Rev. D}\ }\textbf {\bibinfo {volume}
  {103}},\ \bibinfo {pages} {043520} (\bibinfo {year} {2021})},\ \Eprint
  {http://arxiv.org/abs/2007.02943} {arXiv:2007.02943 [astro-ph.CO]}
  \BibitemShut {NoStop}%
\bibitem [{\citenamefont {Abbott}\ \emph
  {et~al.}(2017{\natexlab{a}})\citenamefont {Abbott} \emph
  {et~al.}}]{LIGOScientific:2017vwq}%
  \BibitemOpen
  \bibfield  {author} {\bibinfo {author} {\bibfnamefont {B.~P.}\ \bibnamefont
  {Abbott}} \emph {et~al.} (\bibinfo {collaboration} {LIGO Scientific,
  Virgo}),\ }\href {\doibase 10.1103/PhysRevLett.119.161101} {\bibfield
  {journal} {\bibinfo  {journal} {Phys. Rev. Lett.}\ }\textbf {\bibinfo
  {volume} {119}},\ \bibinfo {pages} {161101} (\bibinfo {year}
  {2017}{\natexlab{a}})},\ \Eprint {http://arxiv.org/abs/1710.05832}
  {arXiv:1710.05832 [gr-qc]} \BibitemShut {NoStop}%
\bibitem [{\citenamefont {Aasi}\ \emph {et~al.}(2015)\citenamefont {Aasi} \emph
  {et~al.}}]{LIGOScientific:2014pky}%
  \BibitemOpen
  \bibfield  {author} {\bibinfo {author} {\bibfnamefont {J.}~\bibnamefont
  {Aasi}} \emph {et~al.} (\bibinfo {collaboration} {LIGO Scientific}),\ }\href
  {\doibase 10.1088/0264-9381/32/7/074001} {\bibfield  {journal} {\bibinfo
  {journal} {Class. Quant. Grav.}\ }\textbf {\bibinfo {volume} {32}},\ \bibinfo
  {pages} {074001} (\bibinfo {year} {2015})},\ \Eprint
  {http://arxiv.org/abs/1411.4547} {arXiv:1411.4547 [gr-qc]} \BibitemShut
  {NoStop}%
\bibitem [{\citenamefont {Acernese}\ \emph {et~al.}(2015)\citenamefont
  {Acernese} \emph {et~al.}}]{VIRGO:2014yos}%
  \BibitemOpen
  \bibfield  {author} {\bibinfo {author} {\bibfnamefont {F.}~\bibnamefont
  {Acernese}} \emph {et~al.} (\bibinfo {collaboration} {VIRGO}),\ }\href
  {\doibase 10.1088/0264-9381/32/2/024001} {\bibfield  {journal} {\bibinfo
  {journal} {Class. Quant. Grav.}\ }\textbf {\bibinfo {volume} {32}},\ \bibinfo
  {pages} {024001} (\bibinfo {year} {2015})},\ \Eprint
  {http://arxiv.org/abs/1408.3978} {arXiv:1408.3978 [gr-qc]} \BibitemShut
  {NoStop}%
\bibitem [{\citenamefont {Abbott}\ \emph
  {et~al.}(2017{\natexlab{b}})\citenamefont {Abbott} \emph
  {et~al.}}]{LIGOScientific:2017adf}%
  \BibitemOpen
  \bibfield  {author} {\bibinfo {author} {\bibfnamefont {B.~P.}\ \bibnamefont
  {Abbott}} \emph {et~al.} (\bibinfo {collaboration} {LIGO Scientific, Virgo,
  1M2H, Dark Energy Camera GW-E, DES, DLT40, Las Cumbres Observatory, VINROUGE,
  MASTER}),\ }\href {\doibase 10.1038/nature24471} {\bibfield  {journal}
  {\bibinfo  {journal} {Nature}\ }\textbf {\bibinfo {volume} {551}},\ \bibinfo
  {pages} {85} (\bibinfo {year} {2017}{\natexlab{b}})},\ \Eprint
  {http://arxiv.org/abs/1710.05835} {arXiv:1710.05835 [astro-ph.CO]}
  \BibitemShut {NoStop}%
\bibitem [{\citenamefont {Soares-Santos}\ \emph {et~al.}(2017)\citenamefont
  {Soares-Santos} \emph {et~al.}}]{DES:2017kbs}%
  \BibitemOpen
  \bibfield  {author} {\bibinfo {author} {\bibfnamefont {M.}~\bibnamefont
  {Soares-Santos}} \emph {et~al.} (\bibinfo {collaboration} {DES, Dark Energy
  Camera GW-EM}),\ }\href {\doibase 10.3847/2041-8213/aa9059} {\bibfield
  {journal} {\bibinfo  {journal} {Astrophys. J. Lett.}\ }\textbf {\bibinfo
  {volume} {848}},\ \bibinfo {pages} {L16} (\bibinfo {year} {2017})},\ \Eprint
  {http://arxiv.org/abs/1710.05459} {arXiv:1710.05459 [astro-ph.HE]}
  \BibitemShut {NoStop}%
\bibitem [{\citenamefont {Abbott}\ \emph
  {et~al.}(2021{\natexlab{a}})\citenamefont {Abbott} \emph
  {et~al.}}]{LIGOScientific:2021aug}%
  \BibitemOpen
  \bibfield  {author} {\bibinfo {author} {\bibfnamefont {R.}~\bibnamefont
  {Abbott}} \emph {et~al.} (\bibinfo {collaboration} {LIGO Scientific, VIRGO,
  KAGRA}),\ }\href@noop {} {\  (\bibinfo {year} {2021}{\natexlab{a}})},\
  \Eprint {http://arxiv.org/abs/2111.03604} {arXiv:2111.03604 [astro-ph.CO]}
  \BibitemShut {NoStop}%
\bibitem [{\citenamefont {Abbott}\ \emph
  {et~al.}(2021{\natexlab{b}})\citenamefont {Abbott} \emph
  {et~al.}}]{LIGOScientific:2021djp}%
  \BibitemOpen
  \bibfield  {author} {\bibinfo {author} {\bibfnamefont {R.}~\bibnamefont
  {Abbott}} \emph {et~al.} (\bibinfo {collaboration} {LIGO Scientific, VIRGO,
  KAGRA}),\ }\href@noop {} {\  (\bibinfo {year} {2021}{\natexlab{b}})},\
  \Eprint {http://arxiv.org/abs/2111.03606} {arXiv:2111.03606 [gr-qc]}
  \BibitemShut {NoStop}%
\bibitem [{\citenamefont {Mastrogiovanni}\ \emph {et~al.}(2021)\citenamefont
  {Mastrogiovanni}, \citenamefont {Leyde}, \citenamefont {Karathanasis},
  \citenamefont {Chassande-Mottin}, \citenamefont {Steer}, \citenamefont
  {Gair}, \citenamefont {Ghosh}, \citenamefont {Gray}, \citenamefont
  {Mukherjee},\ and\ \citenamefont {Rinaldi}}]{Mastrogiovanni:2021wsd}%
  \BibitemOpen
  \bibfield  {author} {\bibinfo {author} {\bibfnamefont {S.}~\bibnamefont
  {Mastrogiovanni}}, \bibinfo {author} {\bibfnamefont {K.}~\bibnamefont
  {Leyde}}, \bibinfo {author} {\bibfnamefont {C.}~\bibnamefont {Karathanasis}},
  \bibinfo {author} {\bibfnamefont {E.}~\bibnamefont {Chassande-Mottin}},
  \bibinfo {author} {\bibfnamefont {D.~A.}\ \bibnamefont {Steer}}, \bibinfo
  {author} {\bibfnamefont {J.}~\bibnamefont {Gair}}, \bibinfo {author}
  {\bibfnamefont {A.}~\bibnamefont {Ghosh}}, \bibinfo {author} {\bibfnamefont
  {R.}~\bibnamefont {Gray}}, \bibinfo {author} {\bibfnamefont {S.}~\bibnamefont
  {Mukherjee}}, \ and\ \bibinfo {author} {\bibfnamefont {S.}~\bibnamefont
  {Rinaldi}},\ }\href {\doibase 10.1103/PhysRevD.104.062009} {\bibfield
  {journal} {\bibinfo  {journal} {Phys. Rev. D}\ }\textbf {\bibinfo {volume}
  {104}},\ \bibinfo {pages} {062009} (\bibinfo {year} {2021})},\ \Eprint
  {http://arxiv.org/abs/2103.14663} {arXiv:2103.14663 [gr-qc]} \BibitemShut
  {NoStop}%
\bibitem [{\citenamefont {Abbott}\ \emph
  {et~al.}(2021{\natexlab{c}})\citenamefont {Abbott} \emph
  {et~al.}}]{LIGOScientific:2019zcs}%
  \BibitemOpen
  \bibfield  {author} {\bibinfo {author} {\bibfnamefont {B.~P.}\ \bibnamefont
  {Abbott}} \emph {et~al.} (\bibinfo {collaboration} {LIGO Scientific,
  Virgo}),\ }\href {\doibase 10.3847/1538-4357/abdcb7} {\bibfield  {journal}
  {\bibinfo  {journal} {Astrophys. J.}\ }\textbf {\bibinfo {volume} {909}},\
  \bibinfo {pages} {218} (\bibinfo {year} {2021}{\natexlab{c}})},\ \Eprint
  {http://arxiv.org/abs/1908.06060} {arXiv:1908.06060 [astro-ph.CO]}
  \BibitemShut {NoStop}%
\bibitem [{\citenamefont {Verde}\ \emph {et~al.}(2019)\citenamefont {Verde},
  \citenamefont {Treu},\ and\ \citenamefont {Riess}}]{Verde:2019ivm}%
  \BibitemOpen
  \bibfield  {author} {\bibinfo {author} {\bibfnamefont {L.}~\bibnamefont
  {Verde}}, \bibinfo {author} {\bibfnamefont {T.}~\bibnamefont {Treu}}, \ and\
  \bibinfo {author} {\bibfnamefont {A.~G.}\ \bibnamefont {Riess}},\ }\href
  {\doibase 10.1038/s41550-019-0902-0} {\bibfield  {journal} {\bibinfo
  {journal} {Nature Astron.}\ }\textbf {\bibinfo {volume} {3}},\ \bibinfo
  {pages} {891} (\bibinfo {year} {2019})},\ \Eprint
  {http://arxiv.org/abs/1907.10625} {arXiv:1907.10625 [astro-ph.CO]}
  \BibitemShut {NoStop}%
\bibitem [{\citenamefont {Aghanim}\ \emph {et~al.}(2020)\citenamefont {Aghanim}
  \emph {et~al.}}]{Planck:2018vyg}%
  \BibitemOpen
  \bibfield  {author} {\bibinfo {author} {\bibfnamefont {N.}~\bibnamefont
  {Aghanim}} \emph {et~al.} (\bibinfo {collaboration} {Planck}),\ }\href
  {\doibase 10.1051/0004-6361/201833910} {\bibfield  {journal} {\bibinfo
  {journal} {Astron. Astrophys.}\ }\textbf {\bibinfo {volume} {641}},\ \bibinfo
  {pages} {A6} (\bibinfo {year} {2020})},\ \bibinfo {note} {[Erratum:
  Astron.Astrophys. 652, C4 (2021)]},\ \Eprint
  {http://arxiv.org/abs/1807.06209} {arXiv:1807.06209 [astro-ph.CO]}
  \BibitemShut {NoStop}%
\bibitem [{\citenamefont {Riess}\ \emph {et~al.}(2021)\citenamefont {Riess}
  \emph {et~al.}}]{Riess:2021jrx}%
  \BibitemOpen
  \bibfield  {author} {\bibinfo {author} {\bibfnamefont {A.~G.}\ \bibnamefont
  {Riess}} \emph {et~al.},\ }\href@noop {} {\  (\bibinfo {year} {2021})},\
  \Eprint {http://arxiv.org/abs/2112.04510} {arXiv:2112.04510 [astro-ph.CO]}
  \BibitemShut {NoStop}%
\bibitem [{\citenamefont {Messenger}\ and\ \citenamefont
  {Read}(2012)}]{Messenger:2011gi}%
  \BibitemOpen
  \bibfield  {author} {\bibinfo {author} {\bibfnamefont {C.}~\bibnamefont
  {Messenger}}\ and\ \bibinfo {author} {\bibfnamefont {J.}~\bibnamefont
  {Read}},\ }\href {\doibase 10.1103/PhysRevLett.108.091101} {\bibfield
  {journal} {\bibinfo  {journal} {Phys. Rev. Lett.}\ }\textbf {\bibinfo
  {volume} {108}},\ \bibinfo {pages} {091101} (\bibinfo {year} {2012})},\
  \Eprint {http://arxiv.org/abs/1107.5725} {arXiv:1107.5725 [gr-qc]}
  \BibitemShut {NoStop}%
\bibitem [{\citenamefont {Flanagan}\ and\ \citenamefont
  {Hinderer}(2008)}]{Flanagan:2007ix}%
  \BibitemOpen
  \bibfield  {author} {\bibinfo {author} {\bibfnamefont {E.~E.}\ \bibnamefont
  {Flanagan}}\ and\ \bibinfo {author} {\bibfnamefont {T.}~\bibnamefont
  {Hinderer}},\ }\href {\doibase 10.1103/PhysRevD.77.021502} {\bibfield
  {journal} {\bibinfo  {journal} {Phys. Rev. D}\ }\textbf {\bibinfo {volume}
  {77}},\ \bibinfo {pages} {021502} (\bibinfo {year} {2008})},\ \Eprint
  {http://arxiv.org/abs/0709.1915} {arXiv:0709.1915 [astro-ph]} \BibitemShut
  {NoStop}%
\bibitem [{\citenamefont {Chatterjee}\ \emph {et~al.}(2021)\citenamefont
  {Chatterjee}, \citenamefont {R.}, \citenamefont {Holder}, \citenamefont
  {Holz}, \citenamefont {Perkins}, \citenamefont {Yagi},\ and\ \citenamefont
  {Yunes}}]{Chatterjee:2021xrm}%
  \BibitemOpen
  \bibfield  {author} {\bibinfo {author} {\bibfnamefont {D.}~\bibnamefont
  {Chatterjee}}, \bibinfo {author} {\bibfnamefont {A.~H.~K.}\ \bibnamefont
  {R.}}, \bibinfo {author} {\bibfnamefont {G.}~\bibnamefont {Holder}}, \bibinfo
  {author} {\bibfnamefont {D.~E.}\ \bibnamefont {Holz}}, \bibinfo {author}
  {\bibfnamefont {S.}~\bibnamefont {Perkins}}, \bibinfo {author} {\bibfnamefont
  {K.}~\bibnamefont {Yagi}}, \ and\ \bibinfo {author} {\bibfnamefont
  {N.}~\bibnamefont {Yunes}},\ }\href {\doibase 10.1103/PhysRevD.104.083528}
  {\bibfield  {journal} {\bibinfo  {journal} {Phys. Rev. D}\ }\textbf {\bibinfo
  {volume} {104}},\ \bibinfo {pages} {083528} (\bibinfo {year} {2021})},\
  \Eprint {http://arxiv.org/abs/2106.06589} {arXiv:2106.06589 [gr-qc]}
  \BibitemShut {NoStop}%
\bibitem [{\citenamefont {Yagi}\ and\ \citenamefont
  {Yunes}(2016)}]{Yagi:2015pkc}%
  \BibitemOpen
  \bibfield  {author} {\bibinfo {author} {\bibfnamefont {K.}~\bibnamefont
  {Yagi}}\ and\ \bibinfo {author} {\bibfnamefont {N.}~\bibnamefont {Yunes}},\
  }\href {\doibase 10.1088/0264-9381/33/13/13LT01} {\bibfield  {journal}
  {\bibinfo  {journal} {Class. Quant. Grav.}\ }\textbf {\bibinfo {volume}
  {33}},\ \bibinfo {pages} {13LT01} (\bibinfo {year} {2016})},\ \Eprint
  {http://arxiv.org/abs/1512.02639} {arXiv:1512.02639 [gr-qc]} \BibitemShut
  {NoStop}%
\bibitem [{\citenamefont {Biswas}\ \emph
  {et~al.}(2021{\natexlab{a}})\citenamefont {Biswas}, \citenamefont {Char},
  \citenamefont {Nandi},\ and\ \citenamefont {Bose}}]{Biswas:2020puz}%
  \BibitemOpen
  \bibfield  {author} {\bibinfo {author} {\bibfnamefont {B.}~\bibnamefont
  {Biswas}}, \bibinfo {author} {\bibfnamefont {P.}~\bibnamefont {Char}},
  \bibinfo {author} {\bibfnamefont {R.}~\bibnamefont {Nandi}}, \ and\ \bibinfo
  {author} {\bibfnamefont {S.}~\bibnamefont {Bose}},\ }\href {\doibase
  10.1103/PhysRevD.103.103015} {\bibfield  {journal} {\bibinfo  {journal}
  {Phys. Rev. D}\ }\textbf {\bibinfo {volume} {103}},\ \bibinfo {pages}
  {103015} (\bibinfo {year} {2021}{\natexlab{a}})},\ \Eprint
  {http://arxiv.org/abs/2008.01582} {arXiv:2008.01582 [astro-ph.HE]}
  \BibitemShut {NoStop}%
\bibitem [{\citenamefont {Biswas}(2021)}]{Biswas:2021yge}%
  \BibitemOpen
  \bibfield  {author} {\bibinfo {author} {\bibfnamefont {B.}~\bibnamefont
  {Biswas}},\ }\href {\doibase 10.3847/1538-4357/ac1c72} {\bibfield  {journal}
  {\bibinfo  {journal} {Astrophys. J.}\ }\textbf {\bibinfo {volume} {921}},\
  \bibinfo {pages} {63} (\bibinfo {year} {2021})},\ \Eprint
  {http://arxiv.org/abs/2105.02886} {arXiv:2105.02886 [astro-ph.HE]}
  \BibitemShut {NoStop}%
\bibitem [{\citenamefont {Biswas}\ \emph
  {et~al.}(2021{\natexlab{b}})\citenamefont {Biswas}, \citenamefont {Nandi},
  \citenamefont {Char}, \citenamefont {Bose},\ and\ \citenamefont
  {Stergioulas}}]{Biswas:2020xna}%
  \BibitemOpen
  \bibfield  {author} {\bibinfo {author} {\bibfnamefont {B.}~\bibnamefont
  {Biswas}}, \bibinfo {author} {\bibfnamefont {R.}~\bibnamefont {Nandi}},
  \bibinfo {author} {\bibfnamefont {P.}~\bibnamefont {Char}}, \bibinfo {author}
  {\bibfnamefont {S.}~\bibnamefont {Bose}}, \ and\ \bibinfo {author}
  {\bibfnamefont {N.}~\bibnamefont {Stergioulas}},\ }\href {\doibase
  10.1093/mnras/stab1383} {\bibfield  {journal} {\bibinfo  {journal} {Mon. Not.
  Roy. Astron. Soc.}\ }\textbf {\bibinfo {volume} {505}},\ \bibinfo {pages}
  {1600} (\bibinfo {year} {2021}{\natexlab{b}})},\ \Eprint
  {http://arxiv.org/abs/2010.02090} {arXiv:2010.02090 [astro-ph.HE]}
  \BibitemShut {NoStop}%
\bibitem [{\citenamefont {Gamba}\ \emph {et~al.}(2020)\citenamefont {Gamba},
  \citenamefont {Read},\ and\ \citenamefont {Wade}}]{Gamba:2019kwu}%
  \BibitemOpen
  \bibfield  {author} {\bibinfo {author} {\bibfnamefont {R.}~\bibnamefont
  {Gamba}}, \bibinfo {author} {\bibfnamefont {J.~S.}\ \bibnamefont {Read}}, \
  and\ \bibinfo {author} {\bibfnamefont {L.~E.}\ \bibnamefont {Wade}},\ }\href
  {\doibase 10.1088/1361-6382/ab5ba4} {\bibfield  {journal} {\bibinfo
  {journal} {Class. Quant. Grav.}\ }\textbf {\bibinfo {volume} {37}},\ \bibinfo
  {pages} {025008} (\bibinfo {year} {2020})},\ \Eprint
  {http://arxiv.org/abs/1902.04616} {arXiv:1902.04616 [gr-qc]} \BibitemShut
  {NoStop}%
\bibitem [{\citenamefont {Biswas}\ \emph {et~al.}(2019)\citenamefont {Biswas},
  \citenamefont {Nandi}, \citenamefont {Char},\ and\ \citenamefont
  {Bose}}]{Biswas:2019ifs}%
  \BibitemOpen
  \bibfield  {author} {\bibinfo {author} {\bibfnamefont {B.}~\bibnamefont
  {Biswas}}, \bibinfo {author} {\bibfnamefont {R.}~\bibnamefont {Nandi}},
  \bibinfo {author} {\bibfnamefont {P.}~\bibnamefont {Char}}, \ and\ \bibinfo
  {author} {\bibfnamefont {S.}~\bibnamefont {Bose}},\ }\href {\doibase
  10.1103/PhysRevD.100.044056} {\bibfield  {journal} {\bibinfo  {journal}
  {Phys. Rev. D}\ }\textbf {\bibinfo {volume} {100}},\ \bibinfo {pages}
  {044056} (\bibinfo {year} {2019})},\ \Eprint
  {http://arxiv.org/abs/1905.00678} {arXiv:1905.00678 [gr-qc]} \BibitemShut
  {NoStop}%
\bibitem [{\citenamefont {{Baym}}\ \emph {et~al.}(1971)\citenamefont {{Baym}},
  \citenamefont {{Pethick}},\ and\ \citenamefont
  {{Sutherland}}}]{1971ApJ...170..299B}%
  \BibitemOpen
  \bibfield  {author} {\bibinfo {author} {\bibfnamefont {G.}~\bibnamefont
  {{Baym}}}, \bibinfo {author} {\bibfnamefont {C.}~\bibnamefont {{Pethick}}}, \
  and\ \bibinfo {author} {\bibfnamefont {P.}~\bibnamefont {{Sutherland}}},\
  }\href {\doibase 10.1086/151216} {\bibfield  {journal} {\bibinfo  {journal}
  {\apj}\ }\textbf {\bibinfo {volume} {170}},\ \bibinfo {pages} {299} (\bibinfo
  {year} {1971})}\BibitemShut {NoStop}%
\bibitem [{\citenamefont {Xie}\ and\ \citenamefont {Li}(2019)}]{Xie:2019sqb}%
  \BibitemOpen
  \bibfield  {author} {\bibinfo {author} {\bibfnamefont {W.-J.}\ \bibnamefont
  {Xie}}\ and\ \bibinfo {author} {\bibfnamefont {B.-A.}\ \bibnamefont {Li}},\
  }\href {\doibase 10.3847/1538-4357/ab3f37} {\bibfield  {journal} {\bibinfo
  {journal} {Astrophys. J.}\ }\textbf {\bibinfo {volume} {883}},\ \bibinfo
  {pages} {174} (\bibinfo {year} {2019})},\ \Eprint
  {http://arxiv.org/abs/1907.10741} {arXiv:1907.10741 [astro-ph.HE]}
  \BibitemShut {NoStop}%
\bibitem [{\citenamefont {Oertel}\ \emph {et~al.}(2017)\citenamefont {Oertel},
  \citenamefont {Hempel}, \citenamefont {Kl\"ahn},\ and\ \citenamefont
  {Typel}}]{Oertel:2016bki}%
  \BibitemOpen
  \bibfield  {author} {\bibinfo {author} {\bibfnamefont {M.}~\bibnamefont
  {Oertel}}, \bibinfo {author} {\bibfnamefont {M.}~\bibnamefont {Hempel}},
  \bibinfo {author} {\bibfnamefont {T.}~\bibnamefont {Kl\"ahn}}, \ and\
  \bibinfo {author} {\bibfnamefont {S.}~\bibnamefont {Typel}},\ }\href
  {\doibase 10.1103/RevModPhys.89.015007} {\bibfield  {journal} {\bibinfo
  {journal} {Rev. Mod. Phys.}\ }\textbf {\bibinfo {volume} {89}},\ \bibinfo
  {pages} {015007} (\bibinfo {year} {2017})},\ \Eprint
  {http://arxiv.org/abs/1610.03361} {arXiv:1610.03361 [astro-ph.HE]}
  \BibitemShut {NoStop}%
\bibitem [{\citenamefont {Read}\ \emph {et~al.}(2009)\citenamefont {Read},
  \citenamefont {Lackey}, \citenamefont {Owen},\ and\ \citenamefont
  {Friedman}}]{Read:2008iy}%
  \BibitemOpen
  \bibfield  {author} {\bibinfo {author} {\bibfnamefont {J.~S.}\ \bibnamefont
  {Read}}, \bibinfo {author} {\bibfnamefont {B.~D.}\ \bibnamefont {Lackey}},
  \bibinfo {author} {\bibfnamefont {B.~J.}\ \bibnamefont {Owen}}, \ and\
  \bibinfo {author} {\bibfnamefont {J.~L.}\ \bibnamefont {Friedman}},\ }\href
  {\doibase 10.1103/PhysRevD.79.124032} {\bibfield  {journal} {\bibinfo
  {journal} {Phys. Rev. D}\ }\textbf {\bibinfo {volume} {79}},\ \bibinfo
  {pages} {124032} (\bibinfo {year} {2009})},\ \Eprint
  {http://arxiv.org/abs/0812.2163} {arXiv:0812.2163 [astro-ph]} \BibitemShut
  {NoStop}%
\bibitem [{\citenamefont {Biswas}\ and\ \citenamefont
  {Datta}(2021)}]{Biswas:2021paf}%
  \BibitemOpen
  \bibfield  {author} {\bibinfo {author} {\bibfnamefont {B.}~\bibnamefont
  {Biswas}}\ and\ \bibinfo {author} {\bibfnamefont {S.}~\bibnamefont {Datta}},\
  }\href@noop {} {\  (\bibinfo {year} {2021})},\ \Eprint
  {http://arxiv.org/abs/2112.10824} {arXiv:2112.10824 [astro-ph.HE]}
  \BibitemShut {NoStop}%
\bibitem [{\citenamefont {Yagi}\ and\ \citenamefont
  {Yunes}(2017)}]{Yagi:2016qmr}%
  \BibitemOpen
  \bibfield  {author} {\bibinfo {author} {\bibfnamefont {K.}~\bibnamefont
  {Yagi}}\ and\ \bibinfo {author} {\bibfnamefont {N.}~\bibnamefont {Yunes}},\
  }\href {\doibase 10.1088/1361-6382/34/1/015006} {\bibfield  {journal}
  {\bibinfo  {journal} {Class. Quant. Grav.}\ }\textbf {\bibinfo {volume}
  {34}},\ \bibinfo {pages} {015006} (\bibinfo {year} {2017})},\ \Eprint
  {http://arxiv.org/abs/1608.06187} {arXiv:1608.06187 [gr-qc]} \BibitemShut
  {NoStop}%
\bibitem [{\citenamefont {Evans}\ \emph {et~al.}(2021)\citenamefont {Evans}
  \emph {et~al.}}]{Evans:2021gyd}%
  \BibitemOpen
  \bibfield  {author} {\bibinfo {author} {\bibfnamefont {M.}~\bibnamefont
  {Evans}} \emph {et~al.},\ }\href@noop {} {\  (\bibinfo {year} {2021})},\
  \Eprint {http://arxiv.org/abs/2109.09882} {arXiv:2109.09882 [astro-ph.IM]}
  \BibitemShut {NoStop}%
\bibitem [{\citenamefont {Maggiore}\ \emph {et~al.}(2020)\citenamefont
  {Maggiore} \emph {et~al.}}]{Maggiore:2019uih}%
  \BibitemOpen
  \bibfield  {author} {\bibinfo {author} {\bibfnamefont {M.}~\bibnamefont
  {Maggiore}} \emph {et~al.},\ }\href {\doibase 10.1088/1475-7516/2020/03/050}
  {\bibfield  {journal} {\bibinfo  {journal} {JCAP}\ }\textbf {\bibinfo
  {volume} {03}},\ \bibinfo {pages} {050} (\bibinfo {year} {2020})},\ \Eprint
  {http://arxiv.org/abs/1912.02622} {arXiv:1912.02622 [astro-ph.CO]}
  \BibitemShut {NoStop}%
\bibitem [{\citenamefont {Fryer}\ \emph {et~al.}(2012)\citenamefont {Fryer},
  \citenamefont {Belczynski}, \citenamefont {Wiktorowicz}, \citenamefont
  {Dominik}, \citenamefont {Kalogera},\ and\ \citenamefont
  {Holz}}]{2012ApJ...749...91F}%
  \BibitemOpen
  \bibfield  {author} {\bibinfo {author} {\bibfnamefont {C.~L.}\ \bibnamefont
  {Fryer}}, \bibinfo {author} {\bibfnamefont {K.}~\bibnamefont {Belczynski}},
  \bibinfo {author} {\bibfnamefont {G.}~\bibnamefont {Wiktorowicz}}, \bibinfo
  {author} {\bibfnamefont {M.}~\bibnamefont {Dominik}}, \bibinfo {author}
  {\bibfnamefont {V.}~\bibnamefont {Kalogera}}, \ and\ \bibinfo {author}
  {\bibfnamefont {D.~E.}\ \bibnamefont {Holz}},\ }\href {\doibase
  10.1088/0004-637X/749/1/91} {\bibfield  {journal} {\bibinfo  {journal}
  {\apj}\ }\textbf {\bibinfo {volume} {749}},\ \bibinfo {eid} {91} (\bibinfo
  {year} {2012})},\ \Eprint {http://arxiv.org/abs/1110.1726} {arXiv:1110.1726
  [astro-ph.SR]} \BibitemShut {NoStop}%
\bibitem [{\citenamefont {Woosley}\ \emph {et~al.}(2020)\citenamefont
  {Woosley}, \citenamefont {Sukhbold},\ and\ \citenamefont
  {Janka}}]{Woosley:2020mze}%
  \BibitemOpen
  \bibfield  {author} {\bibinfo {author} {\bibfnamefont {S.}~\bibnamefont
  {Woosley}}, \bibinfo {author} {\bibfnamefont {T.}~\bibnamefont {Sukhbold}}, \
  and\ \bibinfo {author} {\bibfnamefont {H.~T.}\ \bibnamefont {Janka}},\ }\href
  {\doibase 10.3847/1538-4357/ab8cc1} {\bibfield  {journal} {\bibinfo
  {journal} {Astrophys. J.}\ }\textbf {\bibinfo {volume} {896}},\ \bibinfo
  {pages} {56} (\bibinfo {year} {2020})},\ \Eprint
  {http://arxiv.org/abs/2001.10492} {arXiv:2001.10492 [astro-ph.HE]}
  \BibitemShut {NoStop}%
\bibitem [{\citenamefont {Abbott}\ \emph {et~al.}(2016)\citenamefont {Abbott}
  \emph {et~al.}}]{LIGOScientific:2016hpm}%
  \BibitemOpen
  \bibfield  {author} {\bibinfo {author} {\bibfnamefont {B.~P.}\ \bibnamefont
  {Abbott}} \emph {et~al.} (\bibinfo {collaboration} {LIGO Scientific,
  Virgo}),\ }\href {\doibase 10.3847/2041-8205/832/2/L21} {\bibfield  {journal}
  {\bibinfo  {journal} {Astrophys. J. Lett.}\ }\textbf {\bibinfo {volume}
  {832}},\ \bibinfo {pages} {L21} (\bibinfo {year} {2016})},\ \Eprint
  {http://arxiv.org/abs/1607.07456} {arXiv:1607.07456 [astro-ph.HE]}
  \BibitemShut {NoStop}%
\bibitem [{\citenamefont {Mandel}\ \emph {et~al.}(2019)\citenamefont {Mandel},
  \citenamefont {Farr},\ and\ \citenamefont {Gair}}]{Mandel:2018mve}%
  \BibitemOpen
  \bibfield  {author} {\bibinfo {author} {\bibfnamefont {I.}~\bibnamefont
  {Mandel}}, \bibinfo {author} {\bibfnamefont {W.~M.}\ \bibnamefont {Farr}}, \
  and\ \bibinfo {author} {\bibfnamefont {J.~R.}\ \bibnamefont {Gair}},\ }\href
  {\doibase 10.1093/mnras/stz896} {\bibfield  {journal} {\bibinfo  {journal}
  {Mon. Not. Roy. Astron. Soc.}\ }\textbf {\bibinfo {volume} {486}},\ \bibinfo
  {pages} {1086} (\bibinfo {year} {2019})},\ \Eprint
  {http://arxiv.org/abs/1809.02063} {arXiv:1809.02063 [physics.data-an]}
  \BibitemShut {NoStop}%
\bibitem [{\citenamefont {Abbott}\ \emph
  {et~al.}(2020{\natexlab{a}})\citenamefont {Abbott} \emph
  {et~al.}}]{LIGOScientific:2019eut}%
  \BibitemOpen
  \bibfield  {author} {\bibinfo {author} {\bibfnamefont {B.~P.}\ \bibnamefont
  {Abbott}} \emph {et~al.} (\bibinfo {collaboration} {LIGO Scientific,
  Virgo}),\ }\href {\doibase 10.1088/1361-6382/ab5f7c} {\bibfield  {journal}
  {\bibinfo  {journal} {Class. Quant. Grav.}\ }\textbf {\bibinfo {volume}
  {37}},\ \bibinfo {pages} {045006} (\bibinfo {year} {2020}{\natexlab{a}})},\
  \Eprint {http://arxiv.org/abs/1908.01012} {arXiv:1908.01012 [gr-qc]}
  \BibitemShut {NoStop}%
\bibitem [{\citenamefont {Chatziioannou}(2020)}]{Chatziioannou:2020pqz}%
  \BibitemOpen
  \bibfield  {author} {\bibinfo {author} {\bibfnamefont {K.}~\bibnamefont
  {Chatziioannou}},\ }\href {\doibase 10.1007/s10714-020-02754-3} {\bibfield
  {journal} {\bibinfo  {journal} {Gen. Rel. Grav.}\ }\textbf {\bibinfo {volume}
  {52}},\ \bibinfo {pages} {109} (\bibinfo {year} {2020})},\ \Eprint
  {http://arxiv.org/abs/2006.03168} {arXiv:2006.03168 [gr-qc]} \BibitemShut
  {NoStop}%
\bibitem [{\citenamefont {Ghosh}\ \emph {et~al.}(2023)\citenamefont {Ghosh}
  \emph {et~al.}}]{dark_bns}%
  \BibitemOpen
  \bibfield  {author} {\bibinfo {author} {\bibfnamefont {T.}~\bibnamefont
  {Ghosh}} \emph {et~al.} (\bibinfo {collaboration} {in preparation}),\
  }\href@noop {} {\  (\bibinfo {year} {2023})}\BibitemShut {NoStop}%
\bibitem [{\citenamefont {Buchner}\ \emph {et~al.}(2014)\citenamefont
  {Buchner}, \citenamefont {Georgakakis}, \citenamefont {Nandra}, \citenamefont
  {Hsu}, \citenamefont {Rangel}, \citenamefont {Brightman}, \citenamefont
  {Merloni}, \citenamefont {Salvato}, \citenamefont {Donley},\ and\
  \citenamefont {Kocevski}}]{Buchner:2014nha}%
  \BibitemOpen
  \bibfield  {author} {\bibinfo {author} {\bibfnamefont {J.}~\bibnamefont
  {Buchner}}, \bibinfo {author} {\bibfnamefont {A.}~\bibnamefont
  {Georgakakis}}, \bibinfo {author} {\bibfnamefont {K.}~\bibnamefont {Nandra}},
  \bibinfo {author} {\bibfnamefont {L.}~\bibnamefont {Hsu}}, \bibinfo {author}
  {\bibfnamefont {C.}~\bibnamefont {Rangel}}, \bibinfo {author} {\bibfnamefont
  {M.}~\bibnamefont {Brightman}}, \bibinfo {author} {\bibfnamefont
  {A.}~\bibnamefont {Merloni}}, \bibinfo {author} {\bibfnamefont
  {M.}~\bibnamefont {Salvato}}, \bibinfo {author} {\bibfnamefont
  {J.}~\bibnamefont {Donley}}, \ and\ \bibinfo {author} {\bibfnamefont
  {D.}~\bibnamefont {Kocevski}},\ }\href {\doibase 10.1051/0004-6361/201322971}
  {\bibfield  {journal} {\bibinfo  {journal} {Astron. Astrophys.}\ }\textbf
  {\bibinfo {volume} {564}},\ \bibinfo {pages} {A125} (\bibinfo {year}
  {2014})},\ \Eprint {http://arxiv.org/abs/1402.0004} {arXiv:1402.0004
  [astro-ph.HE]} \BibitemShut {NoStop}%
\bibitem [{\citenamefont {Seabold}\ and\ \citenamefont
  {Perktold}(2010)}]{seabold2010statsmodels}%
  \BibitemOpen
  \bibfield  {author} {\bibinfo {author} {\bibfnamefont {S.}~\bibnamefont
  {Seabold}}\ and\ \bibinfo {author} {\bibfnamefont {J.}~\bibnamefont
  {Perktold}},\ }in\ \href {\doibase 10.25080/Majora-92bf1922-011} {\emph
  {\bibinfo {booktitle} {Proceedings of the 9th Python in Science
  Conference}}}\ (\bibinfo {year} {2010})\ pp.\ \bibinfo {pages} {92 --
  96}\BibitemShut {NoStop}%
\bibitem [{\citenamefont {Abbott}\ \emph
  {et~al.}(2021{\natexlab{d}})\citenamefont {Abbott} \emph
  {et~al.}}]{LIGOScientific:2021psn}%
  \BibitemOpen
  \bibfield  {author} {\bibinfo {author} {\bibfnamefont {R.}~\bibnamefont
  {Abbott}} \emph {et~al.} (\bibinfo {collaboration} {LIGO Scientific, VIRGO,
  KAGRA}),\ }\href@noop {} {\  (\bibinfo {year} {2021}{\natexlab{d}})},\
  \Eprint {http://arxiv.org/abs/2111.03634} {arXiv:2111.03634 [astro-ph.HE]}
  \BibitemShut {NoStop}%
\bibitem [{\citenamefont {Riley}\ \emph {et~al.}(2019)\citenamefont {Riley}
  \emph {et~al.}}]{Riley:2019yda}%
  \BibitemOpen
  \bibfield  {author} {\bibinfo {author} {\bibfnamefont {T.~E.}\ \bibnamefont
  {Riley}} \emph {et~al.},\ }\href {\doibase 10.3847/2041-8213/ab481c}
  {\bibfield  {journal} {\bibinfo  {journal} {Astrophys. J. Lett.}\ }\textbf
  {\bibinfo {volume} {887}},\ \bibinfo {pages} {L21} (\bibinfo {year}
  {2019})},\ \Eprint {http://arxiv.org/abs/1912.05702} {arXiv:1912.05702
  [astro-ph.HE]} \BibitemShut {NoStop}%
\bibitem [{\citenamefont {Miller}\ \emph {et~al.}(2019)\citenamefont {Miller}
  \emph {et~al.}}]{Miller:2019cac}%
  \BibitemOpen
  \bibfield  {author} {\bibinfo {author} {\bibfnamefont {M.~C.}\ \bibnamefont
  {Miller}} \emph {et~al.},\ }\href {\doibase 10.3847/2041-8213/ab50c5}
  {\bibfield  {journal} {\bibinfo  {journal} {Astrophys. J. Lett.}\ }\textbf
  {\bibinfo {volume} {887}},\ \bibinfo {pages} {L24} (\bibinfo {year}
  {2019})},\ \Eprint {http://arxiv.org/abs/1912.05705} {arXiv:1912.05705
  [astro-ph.HE]} \BibitemShut {NoStop}%
\bibitem [{\citenamefont {Riley}\ \emph {et~al.}(2021)\citenamefont {Riley}
  \emph {et~al.}}]{Riley:2021pdl}%
  \BibitemOpen
  \bibfield  {author} {\bibinfo {author} {\bibfnamefont {T.~E.}\ \bibnamefont
  {Riley}} \emph {et~al.},\ }\href {\doibase 10.3847/2041-8213/ac0a81}
  {\bibfield  {journal} {\bibinfo  {journal} {Astrophys. J. Lett.}\ }\textbf
  {\bibinfo {volume} {918}},\ \bibinfo {pages} {L27} (\bibinfo {year}
  {2021})},\ \Eprint {http://arxiv.org/abs/2105.06980} {arXiv:2105.06980
  [astro-ph.HE]} \BibitemShut {NoStop}%
\bibitem [{\citenamefont {Miller}\ \emph {et~al.}(2021)\citenamefont {Miller}
  \emph {et~al.}}]{Miller:2021qha}%
  \BibitemOpen
  \bibfield  {author} {\bibinfo {author} {\bibfnamefont {M.~C.}\ \bibnamefont
  {Miller}} \emph {et~al.},\ }\href {\doibase 10.3847/2041-8213/ac089b}
  {\bibfield  {journal} {\bibinfo  {journal} {Astrophys. J. Lett.}\ }\textbf
  {\bibinfo {volume} {918}},\ \bibinfo {pages} {L28} (\bibinfo {year}
  {2021})},\ \Eprint {http://arxiv.org/abs/2105.06979} {arXiv:2105.06979
  [astro-ph.HE]} \BibitemShut {NoStop}%
\bibitem [{\citenamefont {Abbott}\ \emph {et~al.}(2018)\citenamefont {Abbott}
  \emph {et~al.}}]{LIGOScientific:2018cki}%
  \BibitemOpen
  \bibfield  {author} {\bibinfo {author} {\bibfnamefont {B.~P.}\ \bibnamefont
  {Abbott}} \emph {et~al.} (\bibinfo {collaboration} {LIGO Scientific,
  Virgo}),\ }\href {\doibase 10.1103/PhysRevLett.121.161101} {\bibfield
  {journal} {\bibinfo  {journal} {Phys. Rev. Lett.}\ }\textbf {\bibinfo
  {volume} {121}},\ \bibinfo {pages} {161101} (\bibinfo {year} {2018})},\
  \Eprint {http://arxiv.org/abs/1805.11581} {arXiv:1805.11581 [gr-qc]}
  \BibitemShut {NoStop}%
\bibitem [{\citenamefont {Abbott}\ \emph
  {et~al.}(2017{\natexlab{c}})\citenamefont {Abbott}, \citenamefont {Abbott},
  \citenamefont {Abbott}, \citenamefont {Abernathy}, \citenamefont {Ackley},
  \citenamefont {Adams}, \citenamefont {Addesso}, \citenamefont {Adhikari},
  \citenamefont {Adya}, \citenamefont {Affeldt},\ and\ \citenamefont
  {et~al.}}]{Abbott_2017}%
  \BibitemOpen
  \bibfield  {author} {\bibinfo {author} {\bibfnamefont {B.~P.}\ \bibnamefont
  {Abbott}}, \bibinfo {author} {\bibfnamefont {R.}~\bibnamefont {Abbott}},
  \bibinfo {author} {\bibfnamefont {T.~D.}\ \bibnamefont {Abbott}}, \bibinfo
  {author} {\bibfnamefont {M.~R.}\ \bibnamefont {Abernathy}}, \bibinfo {author}
  {\bibfnamefont {K.}~\bibnamefont {Ackley}}, \bibinfo {author} {\bibfnamefont
  {C.}~\bibnamefont {Adams}}, \bibinfo {author} {\bibfnamefont
  {P.}~\bibnamefont {Addesso}}, \bibinfo {author} {\bibfnamefont {R.~X.}\
  \bibnamefont {Adhikari}}, \bibinfo {author} {\bibfnamefont {V.~B.}\
  \bibnamefont {Adya}}, \bibinfo {author} {\bibfnamefont {C.}~\bibnamefont
  {Affeldt}}, \ and\ \bibinfo {author} {\bibnamefont {et~al.}},\ }\href
  {\doibase 10.1088/1361-6382/aa51f4} {\bibfield  {journal} {\bibinfo
  {journal} {Classical and Quantum Gravity}\ }\textbf {\bibinfo {volume}
  {34}},\ \bibinfo {pages} {044001} (\bibinfo {year}
  {2017}{\natexlab{c}})}\BibitemShut {NoStop}%
\bibitem [{\citenamefont {Dietrich}\ \emph {et~al.}(2019)\citenamefont
  {Dietrich}, \citenamefont {Samajdar}, \citenamefont {Khan}, \citenamefont
  {Johnson-McDaniel}, \citenamefont {Dudi},\ and\ \citenamefont
  {Tichy}}]{Dietrich:2019kaq}%
  \BibitemOpen
  \bibfield  {author} {\bibinfo {author} {\bibfnamefont {T.}~\bibnamefont
  {Dietrich}}, \bibinfo {author} {\bibfnamefont {A.}~\bibnamefont {Samajdar}},
  \bibinfo {author} {\bibfnamefont {S.}~\bibnamefont {Khan}}, \bibinfo {author}
  {\bibfnamefont {N.~K.}\ \bibnamefont {Johnson-McDaniel}}, \bibinfo {author}
  {\bibfnamefont {R.}~\bibnamefont {Dudi}}, \ and\ \bibinfo {author}
  {\bibfnamefont {W.}~\bibnamefont {Tichy}},\ }\href {\doibase
  10.1103/PhysRevD.100.044003} {\bibfield  {journal} {\bibinfo  {journal}
  {Phys. Rev. D}\ }\textbf {\bibinfo {volume} {100}},\ \bibinfo {pages}
  {044003} (\bibinfo {year} {2019})},\ \Eprint
  {http://arxiv.org/abs/1905.06011} {arXiv:1905.06011 [gr-qc]} \BibitemShut
  {NoStop}%
\bibitem [{\citenamefont {Ashton}\ \emph {et~al.}(2019)\citenamefont {Ashton},
  \citenamefont {Hübner}, \citenamefont {Lasky}, \citenamefont {Talbot},
  \citenamefont {Ackley}, \citenamefont {Biscoveanu}, \citenamefont {Chu},
  \citenamefont {Divakarla}, \citenamefont {Easter}, \citenamefont
  {Goncharov},\ and\ \citenamefont {et~al.}}]{Ashton_2019}%
  \BibitemOpen
  \bibfield  {author} {\bibinfo {author} {\bibfnamefont {G.}~\bibnamefont
  {Ashton}}, \bibinfo {author} {\bibfnamefont {M.}~\bibnamefont {Hübner}},
  \bibinfo {author} {\bibfnamefont {P.~D.}\ \bibnamefont {Lasky}}, \bibinfo
  {author} {\bibfnamefont {C.}~\bibnamefont {Talbot}}, \bibinfo {author}
  {\bibfnamefont {K.}~\bibnamefont {Ackley}}, \bibinfo {author} {\bibfnamefont
  {S.}~\bibnamefont {Biscoveanu}}, \bibinfo {author} {\bibfnamefont
  {Q.}~\bibnamefont {Chu}}, \bibinfo {author} {\bibfnamefont {A.}~\bibnamefont
  {Divakarla}}, \bibinfo {author} {\bibfnamefont {P.~J.}\ \bibnamefont
  {Easter}}, \bibinfo {author} {\bibfnamefont {B.}~\bibnamefont {Goncharov}}, \
  and\ \bibinfo {author} {\bibnamefont {et~al.}},\ }\href {\doibase
  10.3847/1538-4365/ab06fc} {\bibfield  {journal} {\bibinfo  {journal} {The
  Astrophysical Journal Supplement Series}\ }\textbf {\bibinfo {volume}
  {241}},\ \bibinfo {pages} {27} (\bibinfo {year} {2019})}\BibitemShut
  {NoStop}%
\bibitem [{\citenamefont {Landry}\ \emph {et~al.}(2020)\citenamefont {Landry},
  \citenamefont {Essick},\ and\ \citenamefont
  {Chatziioannou}}]{Landry:2020vaw}%
  \BibitemOpen
  \bibfield  {author} {\bibinfo {author} {\bibfnamefont {P.}~\bibnamefont
  {Landry}}, \bibinfo {author} {\bibfnamefont {R.}~\bibnamefont {Essick}}, \
  and\ \bibinfo {author} {\bibfnamefont {K.}~\bibnamefont {Chatziioannou}},\
  }\href {\doibase 10.1103/PhysRevD.101.123007} {\bibfield  {journal} {\bibinfo
   {journal} {Phys. Rev. D}\ }\textbf {\bibinfo {volume} {101}},\ \bibinfo
  {pages} {123007} (\bibinfo {year} {2020})},\ \Eprint
  {http://arxiv.org/abs/2003.04880} {arXiv:2003.04880 [astro-ph.HE]}
  \BibitemShut {NoStop}%
\bibitem [{\citenamefont {Cutler}\ and\ \citenamefont
  {Flanagan}(1994)}]{Cutler:1994ys}%
  \BibitemOpen
  \bibfield  {author} {\bibinfo {author} {\bibfnamefont {C.}~\bibnamefont
  {Cutler}}\ and\ \bibinfo {author} {\bibfnamefont {E.~E.}\ \bibnamefont
  {Flanagan}},\ }\href {\doibase 10.1103/PhysRevD.49.2658} {\bibfield
  {journal} {\bibinfo  {journal} {Phys. Rev. D}\ }\textbf {\bibinfo {volume}
  {49}},\ \bibinfo {pages} {2658} (\bibinfo {year} {1994})},\ \Eprint
  {http://arxiv.org/abs/gr-qc/9402014} {arXiv:gr-qc/9402014} \BibitemShut
  {NoStop}%
\bibitem [{\citenamefont {Aasi}\ \emph {et~al.}(2013)\citenamefont {Aasi} \emph
  {et~al.}}]{LIGOScientific:2013yzb}%
  \BibitemOpen
  \bibfield  {author} {\bibinfo {author} {\bibfnamefont {J.}~\bibnamefont
  {Aasi}} \emph {et~al.} (\bibinfo {collaboration} {LIGO Scientific, VIRGO}),\
  }\href {\doibase 10.1103/PhysRevD.88.062001} {\bibfield  {journal} {\bibinfo
  {journal} {Phys. Rev. D}\ }\textbf {\bibinfo {volume} {88}},\ \bibinfo
  {pages} {062001} (\bibinfo {year} {2013})},\ \Eprint
  {http://arxiv.org/abs/1304.1775} {arXiv:1304.1775 [gr-qc]} \BibitemShut
  {NoStop}%
\bibitem [{\citenamefont {Calder\'on~Bustillo}\ \emph
  {et~al.}(2021)\citenamefont {Calder\'on~Bustillo}, \citenamefont {Leong},
  \citenamefont {Dietrich},\ and\ \citenamefont
  {Lasky}}]{CalderonBustillo:2020kcg}%
  \BibitemOpen
  \bibfield  {author} {\bibinfo {author} {\bibfnamefont {J.}~\bibnamefont
  {Calder\'on~Bustillo}}, \bibinfo {author} {\bibfnamefont {S.~H.~W.}\
  \bibnamefont {Leong}}, \bibinfo {author} {\bibfnamefont {T.}~\bibnamefont
  {Dietrich}}, \ and\ \bibinfo {author} {\bibfnamefont {P.~D.}\ \bibnamefont
  {Lasky}},\ }\href {\doibase 10.3847/2041-8213/abf502} {\bibfield  {journal}
  {\bibinfo  {journal} {Astrophys. J. Lett.}\ }\textbf {\bibinfo {volume}
  {912}},\ \bibinfo {pages} {L10} (\bibinfo {year} {2021})},\ \Eprint
  {http://arxiv.org/abs/2006.11525} {arXiv:2006.11525 [gr-qc]} \BibitemShut
  {NoStop}%
\bibitem [{\citenamefont {Adhikari}\ \emph {et~al.}(2021)\citenamefont
  {Adhikari} \emph {et~al.}}]{PREX:2021umo}%
  \BibitemOpen
  \bibfield  {author} {\bibinfo {author} {\bibfnamefont {D.}~\bibnamefont
  {Adhikari}} \emph {et~al.} (\bibinfo {collaboration} {PREX}),\ }\href
  {\doibase 10.1103/PhysRevLett.126.172502} {\bibfield  {journal} {\bibinfo
  {journal} {Phys. Rev. Lett.}\ }\textbf {\bibinfo {volume} {126}},\ \bibinfo
  {pages} {172502} (\bibinfo {year} {2021})},\ \Eprint
  {http://arxiv.org/abs/2102.10767} {arXiv:2102.10767 [nucl-ex]} \BibitemShut
  {NoStop}%
\bibitem [{\citenamefont {Reed}\ \emph {et~al.}(2021)\citenamefont {Reed},
  \citenamefont {Fattoyev}, \citenamefont {Horowitz},\ and\ \citenamefont
  {Piekarewicz}}]{Reed:2021nqk}%
  \BibitemOpen
  \bibfield  {author} {\bibinfo {author} {\bibfnamefont {B.~T.}\ \bibnamefont
  {Reed}}, \bibinfo {author} {\bibfnamefont {F.~J.}\ \bibnamefont {Fattoyev}},
  \bibinfo {author} {\bibfnamefont {C.~J.}\ \bibnamefont {Horowitz}}, \ and\
  \bibinfo {author} {\bibfnamefont {J.}~\bibnamefont {Piekarewicz}},\ }\href
  {\doibase 10.1103/PhysRevLett.126.172503} {\bibfield  {journal} {\bibinfo
  {journal} {Phys. Rev. Lett.}\ }\textbf {\bibinfo {volume} {126}},\ \bibinfo
  {pages} {172503} (\bibinfo {year} {2021})},\ \Eprint
  {http://arxiv.org/abs/2101.03193} {arXiv:2101.03193 [nucl-th]} \BibitemShut
  {NoStop}%
\bibitem [{\citenamefont {Xu}\ \emph {et~al.}(2020)\citenamefont {Xu},
  \citenamefont {Xie},\ and\ \citenamefont {Li}}]{Xu:2020fdc}%
  \BibitemOpen
  \bibfield  {author} {\bibinfo {author} {\bibfnamefont {J.}~\bibnamefont
  {Xu}}, \bibinfo {author} {\bibfnamefont {W.-J.}\ \bibnamefont {Xie}}, \ and\
  \bibinfo {author} {\bibfnamefont {B.-A.}\ \bibnamefont {Li}},\ }\href
  {\doibase 10.1103/PhysRevC.102.044316} {\bibfield  {journal} {\bibinfo
  {journal} {Phys. Rev. C}\ }\textbf {\bibinfo {volume} {102}},\ \bibinfo
  {pages} {044316} (\bibinfo {year} {2020})},\ \Eprint
  {http://arxiv.org/abs/2007.07669} {arXiv:2007.07669 [nucl-th]} \BibitemShut
  {NoStop}%
\bibitem [{\citenamefont {Adhikari}\ \emph {et~al.}(2022)\citenamefont
  {Adhikari} \emph {et~al.}}]{CREX:2022kgg}%
  \BibitemOpen
  \bibfield  {author} {\bibinfo {author} {\bibfnamefont {D.}~\bibnamefont
  {Adhikari}} \emph {et~al.} (\bibinfo {collaboration} {CREX}),\ }\href
  {\doibase 10.1103/PhysRevLett.129.042501} {\bibfield  {journal} {\bibinfo
  {journal} {Phys. Rev. Lett.}\ }\textbf {\bibinfo {volume} {129}},\ \bibinfo
  {pages} {042501} (\bibinfo {year} {2022})},\ \Eprint
  {http://arxiv.org/abs/2205.11593} {arXiv:2205.11593 [nucl-ex]} \BibitemShut
  {NoStop}%
\bibitem [{\citenamefont {Hu}\ \emph {et~al.}(2021)\citenamefont {Hu} \emph
  {et~al.}}]{Hu:2021trw}%
  \BibitemOpen
  \bibfield  {author} {\bibinfo {author} {\bibfnamefont {B.}~\bibnamefont {Hu}}
  \emph {et~al.},\ }\href {\doibase 10.1038/s41567-022-01715-8} {\  (\bibinfo
  {year} {2021}),\ 10.1038/s41567-022-01715-8},\ \Eprint
  {http://arxiv.org/abs/2112.01125} {arXiv:2112.01125 [nucl-th]} \BibitemShut
  {NoStop}%
\bibitem [{\citenamefont {Vi\~nas}\ \emph {et~al.}(2014)\citenamefont
  {Vi\~nas}, \citenamefont {Centelles}, \citenamefont {Roca-Maza},\ and\
  \citenamefont {Warda}}]{Vinas:2013hua}%
  \BibitemOpen
  \bibfield  {author} {\bibinfo {author} {\bibfnamefont {X.}~\bibnamefont
  {Vi\~nas}}, \bibinfo {author} {\bibfnamefont {M.}~\bibnamefont {Centelles}},
  \bibinfo {author} {\bibfnamefont {X.}~\bibnamefont {Roca-Maza}}, \ and\
  \bibinfo {author} {\bibfnamefont {M.}~\bibnamefont {Warda}},\ }\href
  {\doibase 10.1140/epja/i2014-14027-8} {\bibfield  {journal} {\bibinfo
  {journal} {Eur. Phys. J. A}\ }\textbf {\bibinfo {volume} {50}},\ \bibinfo
  {pages} {27} (\bibinfo {year} {2014})},\ \Eprint
  {http://arxiv.org/abs/1308.1008} {arXiv:1308.1008 [nucl-th]} \BibitemShut
  {NoStop}%
\bibitem [{\citenamefont {Tripathy}\ \emph {et~al.}(2020)\citenamefont
  {Tripathy}, \citenamefont {Behera}, \citenamefont {Routray},\ and\
  \citenamefont {Behera}}]{Tripathy:2020yig}%
  \BibitemOpen
  \bibfield  {author} {\bibinfo {author} {\bibfnamefont {S.~K.}\ \bibnamefont
  {Tripathy}}, \bibinfo {author} {\bibfnamefont {D.}~\bibnamefont {Behera}},
  \bibinfo {author} {\bibfnamefont {T.~R.}\ \bibnamefont {Routray}}, \ and\
  \bibinfo {author} {\bibfnamefont {B.}~\bibnamefont {Behera}},\ }\href@noop {}
  {\  (\bibinfo {year} {2020})},\ \Eprint {http://arxiv.org/abs/2009.00427}
  {arXiv:2009.00427 [nucl-th]} \BibitemShut {NoStop}%
\bibitem [{\citenamefont {Alsing}\ \emph {et~al.}(2018)\citenamefont {Alsing},
  \citenamefont {Silva},\ and\ \citenamefont {Berti}}]{Alsing:2017bbc}%
  \BibitemOpen
  \bibfield  {author} {\bibinfo {author} {\bibfnamefont {J.}~\bibnamefont
  {Alsing}}, \bibinfo {author} {\bibfnamefont {H.~O.}\ \bibnamefont {Silva}}, \
  and\ \bibinfo {author} {\bibfnamefont {E.}~\bibnamefont {Berti}},\ }\href
  {\doibase 10.1093/mnras/sty1065} {\bibfield  {journal} {\bibinfo  {journal}
  {Mon. Not. Roy. Astron. Soc.}\ }\textbf {\bibinfo {volume} {478}},\ \bibinfo
  {pages} {1377} (\bibinfo {year} {2018})},\ \Eprint
  {http://arxiv.org/abs/1709.07889} {arXiv:1709.07889 [astro-ph.HE]}
  \BibitemShut {NoStop}%
\bibitem [{\citenamefont {Landry}\ and\ \citenamefont
  {Read}(2021)}]{Landry:2021hvl}%
  \BibitemOpen
  \bibfield  {author} {\bibinfo {author} {\bibfnamefont {P.}~\bibnamefont
  {Landry}}\ and\ \bibinfo {author} {\bibfnamefont {J.~S.}\ \bibnamefont
  {Read}},\ }\href {\doibase 10.3847/2041-8213/ac2f3e} {\bibfield  {journal}
  {\bibinfo  {journal} {Astrophys. J. Lett.}\ }\textbf {\bibinfo {volume}
  {921}},\ \bibinfo {pages} {L25} (\bibinfo {year} {2021})},\ \Eprint
  {http://arxiv.org/abs/2107.04559} {arXiv:2107.04559 [astro-ph.HE]}
  \BibitemShut {NoStop}%
\bibitem [{\citenamefont {Sathyaprakash}\ and\ \citenamefont
  {Schutz}(2009)}]{Sathyaprakash:2009xs}%
  \BibitemOpen
  \bibfield  {author} {\bibinfo {author} {\bibfnamefont {B.~S.}\ \bibnamefont
  {Sathyaprakash}}\ and\ \bibinfo {author} {\bibfnamefont {B.~F.}\ \bibnamefont
  {Schutz}},\ }\href {\doibase 10.12942/lrr-2009-2} {\bibfield  {journal}
  {\bibinfo  {journal} {Living Rev. Rel.}\ }\textbf {\bibinfo {volume} {12}},\
  \bibinfo {pages} {2} (\bibinfo {year} {2009})},\ \Eprint
  {http://arxiv.org/abs/0903.0338} {arXiv:0903.0338 [gr-qc]} \BibitemShut
  {NoStop}%
\bibitem [{\citenamefont {Abbott}\ \emph
  {et~al.}(2020{\natexlab{b}})\citenamefont {Abbott} \emph
  {et~al.}}]{LIGOScientific:2019hgc}%
  \BibitemOpen
  \bibfield  {author} {\bibinfo {author} {\bibfnamefont {B.~P.}\ \bibnamefont
  {Abbott}} \emph {et~al.} (\bibinfo {collaboration} {LIGO Scientific,
  Virgo}),\ }\href {\doibase 10.1088/1361-6382/ab685e} {\bibfield  {journal}
  {\bibinfo  {journal} {Class. Quant. Grav.}\ }\textbf {\bibinfo {volume}
  {37}},\ \bibinfo {pages} {055002} (\bibinfo {year} {2020}{\natexlab{b}})},\
  \Eprint {http://arxiv.org/abs/1908.11170} {arXiv:1908.11170 [gr-qc]}
  \BibitemShut {NoStop}%
\end{thebibliography}%
\end{document}